\documentclass[9pt,cites,graphicx]{article}

\usepackage{epsfig}

\begin{document}

%\title{Bound states of the KRb molecule from $v=0$ to the dissociation limit: coupled channels calculations and simple models}

\title{Ultracold molecules from ultracold atoms: a case study with the KRb molecule}

\author{Paul S. Julienne
\\[3mm]
Joint Quantum Institute, National Institute of Standards and \\Technology and The University of Maryland\\100 Bureau Drive Stop 8423, Gaithersburg, Maryland 20899-8423, USA\\E-mail: paul.julienne@nist.gov\\[1mm]
}

\maketitle

\renewcommand{\thefootnote}{\fnsymbol{footnote}}

\noindent Ultracold collisions of cold atoms or molecules make the bound states of the collision complex formed from the two colliding species accessible for control and manipulation of the cold species or the complex.  Such resonances are best treated by a resonant scattering theory, which in the ultracold domain can take advantage of the properties of the long-range potential and the methods of multichannel quantum defect theory.   Coupled channels calculations on the threshold scattering states and bound states of the $^{40}$K$^{87}$Rb molecule illustrate the ideas and methodology of quantum defect theory using the long-range potential and  also demonstrate the spin properties of the bound states throughout the spectrum.

\section{Introduction}\label{Intro}

The success of cooling and trapping of neutral atoms in the ultracold regime on the order of $\mu$K or less has led to a broad range of advances in multidisciplinary research with quantum degenerate gases of bosons~\cite{Ketterle2002,Cornell2002,Dalfovo1999} or fermions~\cite{Varenna2006,Giorgini2008,Bloch2008}.  Recent work has emphasized strongly correlated many-body physics and reduced dimensional physics~\cite{Bloch2008b,Yurovsky2008}.  The collisional interaction between cold atoms is a crucial factor in determining the properties and stability of an ultracold gas.   These properties can in many cases be precisely controlled using tunable scattering resonances known as Feshbach resonances~\cite{Kohler2006,Chin2008}.  

Ultracold atom research is especially promising with atoms confined in optical lattices, which can provide confinement in one, two, or three dimensions~\cite{Jessen1996,Bloch2005,Greiner2008}.  Three dimensional lattices are comprised of a periodic array of trapping cells and offer the possibility of realizing quantum phase transitions whereby each cell holds an individual atom, or a pair of atoms. Cell trapping depths can be on the order of $\mu$K and intercell spacings on the order of hundreds of nm.  Given the ability to control lattice depth and spacing and the strength of interatomic interactions, lattices of cold atoms or molecules can realize a rich variety of physical systems for a variety of applications.

Molecules are more difficult to cool than atoms, because their much more complex internal structure does not allow simple laser cooling methods~\cite{Phillips1998,Chu1998,Cohentannoudji1998} to be applied to them~\cite{Bahns1996}.  While other methods to cool molecules are being developed, such as molecular beam deceleration~\cite{Bethlem1999}, they have not yet succeeded in reaching the ultracold regime; see the review by Krems~\cite{Krems2008}.  So far, the only successful route to getting ultracold molecules in the $\mu$K domain is to associate two already cold atoms to make a dimer molecule.  The first work in this area, reviewed in References~\cite{Jones2006} and~\cite{Hutson2006}, was carried out by a number of groups that used photoassociation to make molecules in magneto-optical traps at relatively low phase space density and translational temperatures on the order of 100 $\mu$K.  This early work relied on spontaneous emission from excited electronic states, which resulted in a distribution of ground state vibrational and rotational levels and negligible population in the vibrational ground state of the molecule.   

The most effective way to make an ultracold molecular gas with high phase space density is to convert atom pairs in an already cold and dense atomic gas to bound molecular states.  This can be done, for example, by using a time-dependent magnetic field to sweep a Feshbach resonance across threshold from higher to lower energy~\cite{Mies2000}.  Kokkelmans {\it et al.}~\cite{Kokkelmans2001} suggested such a sweep could be combined with optical Raman population transfer to enhance the production of deeply bound molecules in the vibrational ground state.  The vibrationally excited, weakly bound threshold molecules are in most cases not collisionally stable but undergo rapid vibrational relaxation upon collision with an atom or another molecule.  On the other hand, $v=0$ levels would not suffer from collisional vibrational relaxation.   

Jaksch {\it et al.}~\cite{Jaksch2002} suggested that cold ground state $^{87}$Rb$_2$ molecules could be made by photoassociation of pairs of $^{87}$Rb atoms in optical lattice cells to make a weakly bound state, followed by a series of Raman population transfer steps to transfer population stepwise to the $v=0$ ground state.   Upon turning off the lattice these vibrational ground state molecules could be expected to form a molecular Bose-Einstein condensate.   Damski {\it et al.}~\cite{Damski2003} suggested a dipolar superfluid could be made in a similar way by associating K and Rb atoms in optical lattice cells and optically converting them to ground state KRb polar molecules.   The essential step in any of these schemes is the first step of associating two atoms to make a weakly bound molecular state.   Once associated, well-established optical techniques should be capable of transferring the population of the highly vibrationally excited molecular state to the ground state.   

Sage {\it et al.}~\cite{Sage2005} showed that a single stimulated Raman step could be used to transfer population from a weakly bound excited vibrational level to the $v=0$ ground state of the RbCs molecule.  Figure 1 schematically illustrates this kind of process.  The initial molecular state in Ref.~\cite{Sage2005} was formed by spontaneous emission in a low-phase space density gas.  However, it is much better to use magnetoassociation to convert atom pairs in a high phase space density gas to very weakly bound molecules known as Feshbach molecules.  K{\"o}hler  {\it et al.}~\cite{Kohler2006} have extensively reviewed research that use magnetically tunable Feshbach resonances to achieve such association.  Molecules formed this way are just as translationally cold as the atoms that are initially present, and their density is similar to that of the atoms when the transfer efficiency is high.  It is now apparent that the key to making deeply bound ultracold molecules is to take advantage of a tunable Feshbach resonance to pair two unbound atoms into a weakly bound molecular state, which in turn can be converted by optical methods to the more deeply bound state.   

A number of homonulcear alkali dimer Feshbach molecules have been made by magnetoassociation: $^{6}$Li$_2$~\cite{Cubizolles2003,Jochim2003a,Strecker2003},  $^{23}$Na$_2$~\cite{Xu2003}, $^{42}$K$_2$~\cite{Greiner2003}, $^{87}$Rb$_2$~\cite{Durr2004}, and $^{133}$Cs$_2$~\cite{Herbig2003,Mark2005}, including molecular Bose-Einstein condensates made of pairs of fermionic atoms~\cite{Bourdel2004,Jochim2003b,Zwierlein2003,Greiner2003}.  Thalheimer {\it et al.}~\cite{Thalhammer2006} demonstrated high efficiency in converting pairs of $^{87}$Rb atoms in a optical lattice cells to make weakly bound $^{87}$Rb$_2$ Feshbach molecules, which are also trapped in the lattice cells.  Winkler {\it et al.}\cite{Winkler2007} were also able to demonstrate efficient coherent optical transfer of population to a lower energy weakly bound $^{87}$Rb$_2$ vibrational level using the STImulated Raman Adiabatic Passage (STIRAP) technique.  Danzl {\it et al.}~\cite{Danzl2008} succeeded in using the STIRAP method to convert a weakly bound $^{133}$Cs$_2$ Feshbach molecule to a much more deeply bound level with a binding energy of around 1000 cm$^{-1}$.

Ni {\it et al.}~\cite{Ni2008} have succeeded in using magnetoassociation plus STIRAP to convert cold atom pairs in a dense gas of $^{40}$K fermions and $^{87}$Rb bosons at 350 nK to $v=0$ $^{40}$K$^{87}$Rb  ground vibrational state molecules in both the a$^3\Sigma_u^+$ state and the X$^1\Sigma_g^+$ electronic ground state.  Previously Zirbel {\it et al.}~\cite{Zirbel2008} had characterized the magnetoassociation to a Feshbach molecule near $54.6$ mT, and Ospelkaus {\it et al.}~\cite{Ospelkaus2008} demonstrated STIRAP to form a dense, ultracold  molecular gas of levels bound by about $E/h=10$ GHz.  Since formation of $^{40}$K$^{87}$Rb in lattice cells has previously been demonstrated by association of two atoms in a lattice cell~\cite{Ospelkaus2006}, it should be straightforward to make a lattice of $v=0$ $^{40}$K$^{87}$Rb molecules.  Polar molecules in lattices have a number of possible applications, including the realization of exotic condensed matter phases~\cite{Micheli2006,Lewenstein2006,Buchler2007} and quantum computation~\cite{Demille2002}.

Collisions are an essential aspect of understanding and applying ultracold atomic or molecular gases or lattices.  Collisions can be beneficial coherent ones that allow control of the system or destructive ones that cause loss or decoherence of trapped atoms or molecules.  The goal of this paper is to illustrate a number of features of ultracold collisions of atoms and molecules using calculations on the bound and scattering states of the  $^{40}$K$^{87}$Rb molecule as an example system.  We use coupled channels calculations with accurate potential energy curves in calculations that extend from the $E=0$ collision threshold to the energy of the deeply bound $v=0$ ground state vibrational level.  It is clear that near-threshold scattering resonances and bound states play a crucial role in ultracold molecule formation.  In particular, we would like to demonstrate the usefulness of a resonant scattering viewpoint, amplified by the insights of generalized multichannel quantum defect theory based on the long-range potential between the colliding species.  

Section~\ref{RScattering} develops the resonant scattering view of an ultracold collision, whereby a collision allows access to a wide range of bound states of a collision complex.  Section~\ref{CC} describes the scattering channels of the $^{40}$K$^{87}$Rb system, the coupled channels method, and the near-threshold scattering resonances in this system.  Section~\ref{qdt} shows how the properties of the van der Waals long range potential determine the qualitative and semi-quantitative features of the near-threshold bound and scattering states.  Section~\ref{deep} describes the character of the molecular levels as energy decreases below threshold to the domain of ''normal'' molecular levels, including the vibrational ground state level. A final section provides a summary of the results of the paper.

\section{The Resonant Scattering Viewpoint}\label{RScattering}

\subsection{Ultracold Collisions for precise spectroscopy and state control}\label{precisespect}

We will begin with some general considerations that are generic to ultracold atomic or molecular collisions.  First, let us assume that the species A and B, each of which could be an atom or molecule, are prepared in specific internal states $|p\rangle_A$ and $|q\rangle_B$ in an ultracold gas at temperature $T$, where relative collision energies tend to be on the order of $E/k_B\approx 1$ $\mu$K, where $k_B$ is the Boltzmann constant; in other units, 1 $\mu$K corresponds to $E/h=21$ kHz, $E/hc=7.0\times10^{-7}$ cm$^{-1}$, or $E=$ 86 peV.  Consequently, the colliding species are prepared in a very precise energy state relative to the energy scale associated with the ''collision complex'' AB, where the fragment A and B interact strongly when they are close together.  The energy scale associated with AB is on the order of typical chemical bond strengths, say $E=1$ eV, equivalent to $E/h=242$ THz or $E/hc=8066$ cm$^{-1}$.  Thus, the complex AB can be prepared with a precise energy spread that may be only 1 part of $10^{10}$ of its ground state binding energy.

Figure 2 shows a schematic view of an ultracold collision, indicating that the species A and B have internal structure and could be prepared in one of several energy states $E_\alpha=E_p+E_q$ of the pair.  For simplicity, we take the energy of the state that is prepared as the zero of the energy scale, illustrated in the Figure as $E_\alpha=E_1=0$.  Atoms typically have hyperfine and Zeeman spin structure in their ground states, whereas molecules will have rotational and vibrational degrees of freedom as well.  Collision channels $\beta$ with $E_\beta>E>E_\alpha$ are ''closed channels'' at the collision at energy $E$, whereas channels $\beta$ with $E>E_\alpha \ge E_\beta$ are ''open channels.''  Collision products can exit the collision in open channels but not closed ones, due to energy conservation.  If more than two atoms are present in the AB species, reactive channels may also be open.

The long-range potential of the AB complex will generally support a series of bound  levels $E_{n\beta}$ leading up to the dissociation limits at $E_\beta$ corresponding to the various internal states $\beta$ of the separated species.  Such levels are indicated schematically for three channels in Fig. 2.  Levels in closed channels with $E_{n\beta}>E_\alpha$ are quasibound levels that can decay to channel $\alpha$, depending of the presence of a coupling term in the Hamiltonian of the AB complex.  Such levels give rise to scattering resonances when collision energy $E$ is near $E_{n\beta}$.  Such resonances can be tuned or coupled to threshold scattering states by external fields.  Magnetically tunable resonances, such as the 54.6 mT $^{40}$K$^{87}$Rb resonance studied in this paper, can be used to control elastic and inelastic collision rates, and to form weakly bound molecular states by time-dependent manipulation of the magnetic field~\cite{Kohler2006,Chin2008}.  Optically tunable resonances, or photoassociation resonances~\cite{Jones2006}, can be tuned or turned on and off by varying the respective frequency or intensity of the driving electromagnetic radiation. Magnetically or optically tunable resonances are treated by formally equivalent theory.  

If the atoms are assumed to start in a scattering state with $E>0$, Fig. 1 gives examples of optically tunable resonances, namely the one-color photoassociation process at frequency $\nu_1$ or the two-color photoassociation process involving $\nu_1$ and $\nu_2$.   Photoassociation is most naturally treated as a resonant scattering process with a decaying resonance level~\cite{Thorsheim1987,Napolitano1994,Fedichev1996,Bohn1999}.  The theory of threshold resonant scattering of a decaying resonance can be readily extended to tightly confining optical lattices of reduced dimension geometries~\cite{Naidon2006} and can be readily adapted to magnetically tunable or other kinds of resonances.

Extraordinary success has been achieved with ultracold atoms with high resolution spectroscopic probing of the states of the AB molecule starting from the precisely prepared states of the atoms.  The collision complex AB can be prepared in a sharp energy state defined by the spread in energy $k_BT$ of the colliding atoms.  Tunable magnetic or radiofrequency probes can tune states within a few GHz of threshold~\cite{Kohler2006,Chin2008}.  One-color photoassociation is especially successful in probing excited states of the AB complex, while two-color Raman photoassociation is especially useful for probing the ground state~\cite{Jones2006}.  Optical methods are capable of tuning to bound states that are removed even by hundreds of THz from threshold with a prcision determined by the linewidths of the lasers involved.   

Perhaps even more importantly than spectroscopic probing, a time-dependent field can be used to associate an A $+$ B pair to make a stable weakly bound AB complex that does not dissociate back to A $+$ B~\cite{Kohler2006}.  This stable complex can then in turn be coupled to other bound states in the near-threshold domain.  Population can then be transferred coherently to other weakly bound near-threshold states using time dependent magnetic~\cite{Mark2007a,Mark2007b}, radiofrequency~\cite{Lang2007}, or optical fields~\cite{Winkler2007,Zirbel2008}.  Time-dependent optical fields can achieve coherent population transfer to deeply bound states~\cite{Danzl2008,Ni2008}.  We have every reason to expect that these techniques will be extended to new species and other domains of frequency, including microwave, THz, and infrared.

Thus, we see that a sample of ultracold atoms--and presumably now samples of ultracold molecules--can be prepared in specific internal states at a precisely defined energy near $E=0$.  The collision of the prepared species then makes available a large part of the entire bound state spectrum of the collision complex for high resolution probing and coherent population transfer.  Bound states of the complex that can be brought into resonance with the near $E=0$ separated species then serve as ''gateway'' states into the rich spectrum of the complex.  This permits very state-specific control over all degrees of freedom of the complex, electronic, vibrational, rotational, spin, and translation.  By confining the stable species AB in a single trapping cell of an optical lattice, even the energy of relative motion is quantized and even more sharply defined.  Furthermore, a molecule in a lattice cell is protected from collisions with A or B atoms or other AB molecules, thus prolonging its lifetime~\cite{Thalhammer2006}.

\subsection{Importance of the long-range potential}\label{longrange}

Given the sensitivity of the near-threshold bound state spectrum to the properties of the long-range form of the potential $V(R)$ between A and B, much can be gained by trying to understand the states of the molecule AB associated with the long range $V(R)$, which varies with some lead power of $R$ as $1/R^n$.  This contrasts with ''normal chemistry'', where one usually seeks to understand the bound states from the ground state level up to the dissociation limit.  In the ultracold domain, collisions are normally much more understandable and even quantitatively treated by starting with the states of the separated species A and B and the states of their complex AB due to the long range interaction between them.  In this way, one does not need to know the full spectrum of the AB molecule in order to characterize the near-threshold domain quite precisely.  This approach has been extraordinarily successful with ultracold atoms~\cite{Kohler2006,Chin2008}; see Refs.~\cite{Marte2002,Wille2007,Hutson2008} for some examples.

For neutral S-state atoms, the long-range potential has the van der Waals form with $n=6$.  Molecules can have dipole or quadrupole moments, corresponding to $n =$ 3 and 5 for two dipoles or two quadrupoles respectively.  These potentials are anisotropic, but have vanishing diagonal matrix elements for $s$-wave interactions in free space.  In the absence of an external field which breaks the symmetry of free space, an isolated polar molecule in a definite state of total angular momentum has a vanishing dipole moment.  However, a strong electric field can induce a dipole moment.

It is convenient to introduce a characteristic length and energy scale associated with the long-range potential.  For this purpose, we use the scale length defined by Gribakin and Flambaum for a potential varying as $-C_n/R^n$~\cite{Gribakin1993},
\begin{equation}
 \bar{a}(n) = \cos\left( \frac{\pi}{n-2}\right )\left ( \frac{2\mu C_n}{\hbar^2(n-2)^2}\right)^\frac{1}{n-2}
 \Gamma\left ( \frac{n-3}{n-2}\right )/ \Gamma \left(\frac{n-1}{n-2} \right ) \,,
 \label{eq:bara}
\end{equation}
where $\mu$ is the reduced mass of the AB pair, $\Gamma$ is the Gamma function, and $\hbar$ is Planck's constant divided by $2\pi$.  This length defines a corresponding energy scale
\begin{equation}
  \bar{E}(n) = \frac{\hbar^2}{2 \mu \bar{a}(n)^2} \,.
\end{equation}
For a van der Waals potential with $n=6$, this simplifies to $\bar{a}=0.477989 (2\mu C_6/\hbar^2)^{1/4}$.  Jones {\it et al.}~\cite{Jones2006} and Chin {\it et al.}~\cite{Chin2008} review the properties of the van der Waals potential relevant to ultracold physics, and Friedrich and Trost~\cite{Friedrich2004} adapt semiclassical theory to obtain the threshold properties for $n=6$ and other cases of $n$.

\subsection{Ultracold resonant scattering theory}\label{resscatt}

Since molecules have more complex internal structure than atoms, subject to electric as well as magnetic and electromagnetic field control, the collisions of cold molecules should also have numerous scattering resonances; for example, see the work of Refs.~\cite{Bohn2003,Avdeenkov2003,Avdeenkov2004,Aldegunde2008}.  Consequently, the theory of resonant scattering is both necessary and useful for understanding cold molecular collisions.  As noted earlier, resonant scattering theory has been extensively developed for photoassociative collisions, which represent optical Feshbach resonances~\cite{Thorsheim1987,Napolitano1994,Fedichev1996,Bohn1999,Naidon2006,Machholm2001,Ciurylo2005}.    

A particularly insightful set of articles on a resonant scattering viewpoint of molecule association and chemical reactions has been given by Mies~\cite{Mies1969a,Mies1969b}.  He considered the role of molecular resonances in the association of two atomic or molecular fragments A and B to form an AB molecule.  He considered both the case of radiative association, which is formally equivalent to the resonant scattering theory of molecule formation by photoassociation, and the case of collisonal association, where a third body deactivates a resonant state of the complex to stabilize it.  In the latter case, the decay if the resonance is simulated by a complex energy with an imaginary term, just as spontaneous emission is represented in radiative association.   The formalism is useful, since it gives a general $S$-matrix resonant scattering theory of association and resonant-enhanced reactions, bounded by the unitarity limit of the $S$-matrix, even when there is a complex set of overlapping resonances.  While the Mies theory was developed for a high temperature gas, where $k_B T$ is large compared to the spacing between the dense set of molecular resonances, there is no reason why the formalism can not carry over directly to the ultracold case, if the threshold properties of the $S$-matrix are incorporated into the theory.  This will be especially useful if the insights of generalized multichannel quantum defect theory~\cite{Mies1984a,Mies1984b} are brought to bear in the ultracold regime~\cite{Mies2000,Julienne1989,Gao2005,Julienne2006}.

The thermally averaged expression from Mies~\cite{Mies1969a,Mies1969b} for the inelastic collision rate coefficient $K_\mathrm{in}(T)$ has a very instructive general form.   Assume species A and B are prepared in channel $\alpha$ in a gas described by a Maxwellian thermal distribution of collision energies at temperature $T$.  Then
\begin{equation}
  K_\mathrm{in}(T) = \frac{1}{Q_T} \frac{k_B T}{h} f_D(T) \,,
  \label{eq:MiesK}
\end{equation}
where the dimensionless dynamical factor $f_D(T)$ is
\begin{equation}
   f_D(T) = \sum_{\ell m_\ell}  \left ( 1 - |S_{\alpha\alpha}(\ell m_\ell)|^2 \right ) e^{-E/(k_B T)} dE/(k_B T) \,.
   \label{eq:MiesfD}
\end{equation}
Here $Q_T$ is the translational partition function, $1/Q_T=(2\pi\mu k_B T/h^2)^{-3/2}=\Lambda_T^3$ where $\Lambda_T$ is the thermal de Broglie wavelength for relative motion of A and B.  The sum defining $f_D(T)$ runs over the contributing partial waves $\ell$ and the $2\ell+1$ projections $m_\ell$ of $\ell$.  Inelastic collisions are those that remove A and B from channel $\alpha \ell m_\ell$, such as loss to a different channel $\beta \ell_\beta m_{\ell \beta}$, which could represent removal of the resonant state through decay, relaxation, or reaction.   If removal of A and B results in formation of an AB molecule, the removal rate is the same as the association rate.  Note that in time-independent scattering theory, production of a resonant quasibound state that only decays back to the entrance channel is not a molecule formation process that can be represented by an $S$-matrix element, since the quasibound state does not persist into the asymptotic domain.  However, if the quasibound resonant state is irreversibly removed to a loss  ''channel'' so that it can not decay back to the entrance channel, then this process can be represented by $S$-matrix scattering theory.  Thus, $S$-matrix resonant scattering theory can be used to represent two-color photoassociation to bound states of the ground state potential, as long as such states have some ''decay width''~\cite{Bohn1999,Julienne1998}.

The removal rate of the density $n_A$ of species A or density $n_B$ of species B is $\dot{n}_A = \dot{n}_B= -K_\mathrm{in} n_A n_B$ (we assume nonidentical species; otherwise identical particle properties would need to be taken into account).  We thus see that the respective removal rates $K_\mathrm{in}n_B$ and $K_\mathrm{in}n_A$ of species A and B are proportional to $n_B \Lambda_T^3$ and $n_A \Lambda_T^3$.  One thus finds that the removal rate of a species is proportional to the dimensionless phase space density  of its collision partner.  The removal rate is also proportional to a thermal factor $k_B T/h$, which sets a basic time scale for the dynamics (e.g., 21 kHz at 1 $\mu$K).    The only part that depends on the specific collision dynamics of the particular system in encapsulated in the dimensionless dynamical factor $\left ( 1 - |S_{\alpha\alpha}(\ell m_\ell)|^2 \right )$ that describes the loss of flux from the $\alpha \ell m_\ell$ entrance channel.  In the case of elastic scattering this factor is replaced by $\left | 1 - S_{\alpha\alpha}(\ell m) \right |^2$ in a corresponding definition of a dynamical $f_D(T)$ expression for elastic scattering.  

The structure of the $S$-matrix elements in the expression for the $f_D$ term in the Mies theory is determined by a set of scattering resonances.  However, the dimensionless $f_D(T)$ factor for inelastic collisions is bounded by the unitarity property of the $S$-matrix so that the contribution from each $\ell m_\ell$ term in the sum has a maximum value of unity.  Thus, if $\ell_\mathrm{max}$ partial waves contribute to the sum, then the bounds on $f_D(T)$ are $0 \le f_D(T) \le (\ell_\mathrm{max}+1)^2$, so that $0 \le f_D \le 1$ for $s$-waves.  The analogous $f_D$ factor for elastic collisions has an upper  bound due to unitarity that is 4 times larger, $4(\ell_\mathrm{max}+1)^2$.  

For $s$-waves in the $E \to 0$ threshold limit, $S = [1- ik(a-ib)]/[1+ik(a-ib)] \approx \exp{[-2ik(a-ib)]}$ is represented by a complex scattering length $a-ib$, where $\hbar k$ is the relative collision momentum, $b$ is nonnegative, and the condition $k|a-ib| \ll 1$ applies.  In this limit, $1-|S|^2= 4kb$, and $K_\mathrm{in}$ reduces to the usual threshold law expression, 
\begin{equation}
 K_\mathrm{in}= 2 (h/\mu) b = 0.84\times 10^{-10} \frac{b[\mathrm{au}]}{\mu[\mathrm{amu}]}\,\,\mathrm{cm}^3/\mathrm{s}\,.
\label{eq:Kin}
\end{equation}
Here $b[\mathrm{au}]$ is in atomic units and $\mu[\mathrm{amu}]$ is in atomic mass units.
Similarly, the $E \to 0$ $s$-wave elastic scattering cross section reduces to the standard form $4\pi (a^2+b^2)$.  Given that $k|a-ib| \ll 1$, the $f_D$ factor for elastic or inelastic collisions remains much less than its upper bound with a value determined by $a$, $b$ and $T$.  Threshold resonance structure can also modify the value of $f_D(T)$, requiring the use of full energy-dependent resonant scattering expressions~\cite{Bohn1999,Naidon2006}.

The expression for $K_\mathrm{in}$ in Eq.~(\ref{eq:MiesK}) is based on very general thermodynamic and dynamical considerations at equilibrium and applies to atomic or molecular collisions.  We may expect Eq.~(\ref{eq:MiesK}) to give a guide to the rate of molecular association even in more general nonequilibrium cases.  It is gratifying that the rate of association of a given species A is proportional to the phase space density of its collision partner, as in semi-empirical treatments of atom association to a Feshbach molecule with time-dependent fields~\cite{Zirbel2008,Hodby2005}.  This is to be expected for a fast process at constant entropy.  Also, time-dependent processes would not be expected to beat the fundamental upper bound to $f_D$ set by the unitarity limit of the time-independent $S$-matrix.  If the dimensionless phase space and dynamical factors in the association rate are both less than unity, as they would be for a Maxwellian gas with an $s$-wave association process in the threshold law domain, then the $k_B T/h$ factor in Eq.~(\ref{eq:MiesK}) sets a limiting upper bound to the rate (and a lower bound to the time scale) for the association of A and B pairs to form AB molecules.

Using the actual threshold resonance form for the $S$-matrix as a function of collision energy~\cite{Bohn1999,Naidon2006} would allow one to work out the general case of resonant elastic and inelastic cross sections and rate constants, including the effect of multiple or overlapping resonances if they are present.  See Machholm {\it et al.}~\cite{Machholm2001}, who develop a coupled channels $S$-matrix resonant scattering theory for decaying resonances, including the effect of overlapping resonances, using a form similar to the Mies theory.  They also develop approximations for describing isolated resonances, using the factorization of matrix elements associated with quantum defect theory.  Coupled channels resonant scattering theory, in its various numerical or approximate analytical representations, should prove to be a very powerful tool for characterizing ultracold molecular collisions, especially if it can take advantage of the long range properties of the interaction potentials to characterize near-threshold states.

\section{Multichannel Scattering of $^{40}$K$^{87}$Rb}\label{CC}

Since ultracold ground state $^{40}$K$^{87}$Rb polar molecules have now been made using resonant association followed by optical population transfer, we will concentrate on using this system to illustrate the features of this process.  We calculate the bound and scattering states of $^{40}$K$^{87}$Rb all the way from threshold to the ground state, including the effect of the long range potentials in determining the states near threshold.  Fortunately, an excellent set of adiabatic Born-Oppenheimer electronic potentials is available for the X$^1\Sigma^+$ and a$^3\Sigma^+$ states of this species, based on the high resolution spectroscopic analysis of Pashov {\it et al.}~\cite{Pashov2007}.  These are the potentials needed to describe the interactions of two ground state atoms.  Using these potentials, and standard representations of the full Hamiltonian of the interactions, including the angular momenta from electron and nuclear spins and internuclear axis rotation, we have carried out full coupled channels calculations~\cite{Kohler2006,Chin2008,Mies2000,Hutson2008,Stoof1988} of the bound and scattering states of this species.

Figures 3 and 4 show the Zeeman substructure of the $^{40}$K and $^{87}$Rb $^2$S ground state atoms as a function of magnetic field.  Each has an electron spin quantum number of $S=1/2$ and respective nuclear spin quantum numbers of $I=$ 4 and $3/2$.  Thus, $^{40}$K  is a composite fermion and $^{87}$Rb is a composite boson, and the $^{40}$K$^{87}$Rb molecule is a fermion.  The figures show how the two zero-field hyperfine levels with total angular momentum $F=I-1/2$ and $F=I+1/2$ split with increasing magnetic field strength $B$.  We use an alphabetical notation to designate each level of the Zeeman manifold by an italic Roman letter $a,b,c\ldots$, starting with the lowest energy state at each $B$ and increasing in order of energy.  In the Figures, the zero of energy is that of the spinless (nonrelativistic) atom.  Thus, the lowest energy hyperfine level at $B=0$ has an energy of $-4E_\mathrm{hf}(^{40}\mathrm{K})/9$ for $^{40}$K and  $-5E_\mathrm{hf}(^{87}\mathrm{Rb})/8$ for $^{87}$Rb, where $E_\mathrm{hf}(^{40}\mathrm{K})/h=1.285790$ GHz  and $E_\mathrm{hf}(^{87}\mathrm{Rb})/h=6.8346826$ GHz are the hyperfine splittings of the two atoms~\cite{Arimondo1977}.

Figure 5 shows the adiabatic Born-Oppenheimer potentials for the X$^1\Sigma^+$ and a$^3\Sigma^+$  electronic states that correlate with the two ground state separated atoms~\cite{Pashov2007}.  The inset to the figure shows the adiabatic potentials (i.e., those that diagonalize the full electronic plus spin Hamiltonian) on an expanded scale showing the long range region.  Since the atoms are $S$-state atoms, all of the long range potentials have the same van der Waals $C_6$ coefficient, which has a value of 4299.51 E$_\mathrm{h}$ a$_0^6$~\cite{Pashov2007}, where E$_\mathrm{h} = 4.359744\times 10^{-18}$ J and a$_0=0.05291772$ nm.  The characteristic van der Waals length for the long range potential is $\bar{a}=68.8$ a$_0$ and the characteristic energy is $\bar{E}/h=13.9$ MHz.

Since we are concerned with $s$-wave resonances in the $aa$ channel, where both atoms are in their lowest energy state, it is necessary to consider all of the $s$-wave states with projection quantum number $M_\mathrm{tot}=-9/2+1=-7/2$, where the $M_\mathrm{tot}$ quantum number also includes the projection $m_\ell$ of partial wave ($m_\ell$ is trivially zero for an $s$-wave).  The projection quantum number is a conserved quantum number at finite field, so that only states with the same value of $M_\mathrm{tot}$ are coupled through terms in the Hamiltonian.  While it is possible to include other partial waves in the expansion basis that are coupled to $s$-waves through anisotropic spin-dependent terms in the Hamiltonian~\cite{Hutson2008}, such coupling terms are small and have a small effect here and need not be included (but see below for the effect of coupling $d$-waves to $s$-waves). 

Figure 6 illustrates the 12 $s$-wave channels that have $M_\mathrm{tot}=-7/2$.  The dotted zero-field curves are the same as in Figure 5, but these split into 12 different curves at finite $B$.  These channels separate themselves into four different groups A, B, C, and D associated with the  four different zero field hyperfine separated atom limits, as described in the caption.  Letting each channel be labeled by the quantum numbers $\beta={ij}$, where $ij$ represents the alphabetic label of the two atoms, with the $^{40}$K label first, the coupled channels expansion of the wave function for atoms initially prepared in channel $\alpha=aa$ is 
\begin{equation}
  \Psi_\alpha(R,E) = \sum_\beta f_{\beta \alpha}(R,E) | \beta \rangle /R \,.
  \label{eq:CC}
\end{equation}
The coupled channels Schr{\"o}dinger equation then determines the solutions with scattering boundary conditions for $E > E_\alpha$ or the discrete set of bound state solutions with bound state boundary conditions for $E < E_\alpha$.  The bound state solutions are found with the discrete variable method described in Ref.~\cite{Hutson2008}, whereas a standard propagator method is used for the scattering solutions.

Figure 7 shows the results of coupled channels scattering and bound state calculations with the 12 $s$-wave basis functions with $M_\mathrm{tot}=-7/2$ in the expansion.    The positions of the four scattering length singularities are in excellent agreement with the measured positions~\cite{Klempt2007,Ferlaino2006a,Ferlaino2006b}.  The resonance near 54.6 mT is well represented near its singularity by the standard resonance formula~\cite{Kohler2006,Chin2008}
\begin{equation}
 a(B) = a_\mathrm{bg} \left ( 1 - \frac{\Delta}{B-B_0} \right ) \,,
 \label{eq:aB}
\end{equation}
where the background scattering length is $a_\mathrm{bg}=-190.6$ a$_0$, the resonance width $\Delta = -0.3103$ mT, and the resonance position is $B_0=54.6937$ mT, compared to measured positions of $54.69$ mT~\cite{Klempt2007} and $54.67$ mT~\cite{Ferlaino2006b}.

The bound levels in Fig. 7 are labeled by the index $\beta$ of the dominant spin component in the coupled channels expansion in Eq. (\ref{eq:CC}) and by the vibrational quantum number $n$ counting down from $n=-1$, which designates the last bound state below the dissociation limit at $E=E_\beta$.  Thus, the line near $-0.4$ GHz parallel to the $E=0$ axis is the last $n=-1$ bound state in the $aa$ entrance channel, whereas the bound state that crosses threshold to make the 54.69 mT resonance is the next to last $n=-2$ bound state of the $rb$ channel, which is associated with the group B of Figure 6.  The two $n=-3$ resonances in the Figure are associated with the highest energy group D.

Figure 8 shows an expanded view of Figure 7 near the 54.69 mT resonance.   When $d$-waves with $\ell=2$, $m_\ell=0$ are added to the basis set, the weak coupling between the $s$- and $d$-waves due to the spin-dipolar interaction shifts $B_0$ from 54.694 mT to 54.687 mT (not shown).  In addition, a new narrow resonance appears on the shoulder of the 54.69 mT resonance at 54.8305 mT with a width of only 4.6 $\mu$T.  The narrow resonance has not been observed.  It is due to the entrance channel $s$-wave being coupled to a bound state of $d$-symmetry.  The bound state is a $\beta(n \ell m_\ell)=rf(-3d0)$ resonance with the same spin and $n$ character as the $rf(-3)$ $s$-wave resonance in Fig. 7, but with $\ell=2$ units of axis rotation angular momentum.  To a good approximation its energy can be obtained by adding the $d$-wave rotational energy for the $n=-3$ level of the $^3\Sigma^+$ potential to the energy of the $s$-wave $rf(-3)$ level.  Finally, Figure 8 shows the universal energy that is derived from the scattering length~\cite{Kohler2006},
\begin{equation}
  E = -\frac{\hbar^2}{2\mu a(B)^2} \,.
  \label{eq:aU}
  \end{equation}
 This universal formula only applies very close to resonance where $a(B) \gg \bar{a}$.  
 
\section{The Quantum Defect Approach with a Long-Range Potential}\label{qdt}

While coupled channels calculations have proven to be an excellent and reliable tool for understanding the near threshold domain of bound and scattering states of ultracold atoms, they are very computer intensive and give a ''black box'' representation of the physics.  Therefore, it is also very desirable to have alternative methods for analysis, understanding, and developing approximations.  In this regard, the general form of multichannel quantum defect theory (MQDT) provides a very powerful set of tools and concepts for the ultracold domain, which takes advantage of the separation between the short-range and long-range aspects of the collision, and exploits the analytic properties of the wavefunction as a function of interatomic separation $R$ and energy $E$.   The concepts of MQDT are implicit in the accumulated phase method pioneered by the Eindhoven group for characterizing ground state interactions of cold atoms~\cite{Tsai1997,vanAbeelen1997,vanAbeelen1999,Vogels1998} and are explicitly developed for neutral atoms in Refs~\cite{Julienne1989,Gao2005,Julienne2006,Mies2000b,Raoult2004} and for ion-atom collisions in Ref.~\cite{Idziaszek2008}.  Gao has developed analytic MQDT for  potentials having the long-range form $-C_n/R^n$~\cite{Gao1998,Gao1998b,Gao2000,Gao2001}.  We will show here several specific ways where MQDT theory has been applied or still needs to be developed for application to the ultracold domain, especially when implemented using the form of the long range potential, 

\subsection{The near-threshold spectrum}

There is a close relation between the near-threshold bound state spectrum and the near-threshold scattering properties of ultracold collisions.  Not only is there a general relation between the scattering length and the last bound state binding energy, as illustrated by the limiting expression in Eq.~(\ref{eq:aU}), but the number and properties of Feshbach resonance states are related to the density and tuning properties of near-threshold bound states.  In view of the importance of the near threshold spectrum for precision spectroscopic probing in Section~\ref{precisespect} and collision  dynamics in Section~\ref{resscatt}, simple ways for understanding the spectrum is desirable.

Gao~\cite{Gao2000,Gao2001} has worked out analytic bound state properties for the van der Waals potential for $n=6$.  We will characterize the long-range $n=6$ potential by the length parameter $\bar{a}$ instead of the $\beta=2.092099 \bar{a}$ parameter used by Gao, since the formulas of the theory are simpler when $\bar{a}$ is used.  The spectrum is completely determined for all partial waves $\ell$ by specifying only three parameters: the reduced mass $\mu$, the $C_6$ constant, and the short range QDT $K_c$-matrix element, or equivalently, the $s$-wave scattering length in units of $\bar{a}$ that uniquely determines the value of $K_c$.  The value of $K_c$ is related to the ''quantum defect'' for the vibrational levels of the potential

Figure 9 shows the spectrum of the last two bound states below threshold for the $^{40}$K$^{87}$Rb molecule for different partial waves $\ell$.    The Figure shows several specific cases.  The blue lines show the spectrum of bound states that one gets for the special case that the $s$-wave scattering length $a=\infty$.  This corresponds to having an $s$-wave bound state at $E=0$.   The Gao theory shows in this case for a pure $-C_6/R^6$ potential that all partial waves with $\ell$ divisible by 4 also have a bound state at $E=0$, as for example, the $g$-wave in Fig. 9.  In the $a=\infty$ case the first $s$-wave level with $E<0$, labeled by $n=-1$ in the Figure, lies at $-36.1\bar{E}=-0.503$ GHz, and the next $n=-2$ level lies at $-249\bar{E}=-3.47$ GHz.  Varying the scattering length over its full range between $-\infty$ and $+\infty$ will produce one $n=-1$ and one $n=-2$ level in the ''bins'' demarked by these values.  Figure 9 also shows the actual energy levels for the X$^1\Sigma^+$ and a$^3\Sigma^+$ potentials, relative to their nonrelativistic separated atom limit at $V(\infty)=0$.  These are near the bottom of their ''bins'',  due to their respective negative scattering lengths of $-111.8$ a$_0$ and $-216.2$ a$_0$~\cite{Pashov2007}.  The actual binding energies for the $n=-1$ and $n=-2$ levels are respectively about 1 percent and 2 percent larger than the same level calculated for a pure van der Waals potential with the same scattering length.  This small difference indicates the small effect of other terms in the potential that contribute to the binding energy of near threshold levels.  It also shows that the van der Waals theory alone is a very good approximate theory, although real potentials need to be used in fitting binding energy data that is more accurate than 1 percent~\cite{Kitagawa2007}.

The figure also shows the bound states for other partial waves.  The energies of these levels can be approximated by adding the rotational energy $\langle n,s |\hbar^2\ell(2\ell+1)/(2\mu R^2) | n,s\rangle$ to the energy of the $s$-wave $|n,s\rangle$ level.  In the angular-momemtum insensitive version of the long range quantum defect theory~\cite{Gao2000,Gao2001}, the energies of the  levels with $\ell \ge1$ are also determined from a knowledge of the $s$-wave scattering  length alone (plus $\mu$ and $C_6$, of course).  However, there will always be a small error since the short range form of the potential never corresponds to $1/R^6$ all the way to $R=0$ but has a finite depth and inner turning point.   For any given entrance channel in the multichannel problem, the near-threshold bound states can be calculated to a good approximation by using the scattering length for that channel to get the $K_c$ matrix for that channel.  

Figure 10 shows the vibrational wave function $f_n(R,E_n)$ for the last three $s$-wave bound states of the $^3\Sigma^+$ potential with energy $E_n$ for $n=-1$, $-2$, and $-3$.  The wave functions show distinct differences for distances on the order of $\bar{a}$.   However,  the three wave functions are remarkably similar in shape at small $R$, where the potential is very deep and the local de Broglie wavelength, which governs the spacing between oscillations, is small compared to $\bar{a}$.  These three unit-normalized wave functions have the same phase at small $R$ and differ only in amplitude.  In fact, these wave functions would have almost identical energy-insensitive amplitudes at short range if they were given a ''quantum defect'' normalization per unit energy by multiplying $f_n(R,E_n)$ by $|\partial n / \partial E|^{1/2}$ for level $n$, where $n$ is viewed as a continuous function of $E$ that takes on integer values at an eigenvalue $E=E_n$ and $|\partial E/\partial n|$ is the mean vibrational spacing between levels at level $n$~\cite{Mies1984a,Mies1984b,Julienne1989} (alternatively $|\partial n / \partial E|$ is the density of states per unit energy).  The wave function can be understood using its energy-normalized semiclassical JWKB form in the short-range classical region, namely $C \sin{\beta(R)}/\sqrt{k(R,E)}$, where $\beta(R)$ is the phase, $\hbar^2 k(R,E)^2 = 2\mu(E-V(R))$ determines the ''local'' momentum wave number $k(R,E)$, and $C=[2\mu/(\pi \hbar^2)]^{1/2}$ is a constant.  In the deep part of the potential, where $E-V(R)  \gg \bar{E}$ is very large and $k(R,E) \approx k(R,0)$, the semiclassical phase and amplitude are not at all sensitive to the value of $E \approx 0$.  This insensitivity leads to the basic concept of MQDT or the accumulated phase method  that the short range physics, such as wave function phases or coupling matrix elements due to short range interactions, can be quite adequately represented by energy-insensitive quantities~\cite{Mies1984a,Mies1984b,Julienne1989}.

\subsection{Feshbach resonance properties}\label{FeshProperties}

The simple van der Waals form of MQDT also gives an excellent way to represent near-threshold scattering and bound state properties of Feshbach resonances  In addition to $\bar{a}$ and $\bar{E}$, the only other quantities that are needed in the near-threshold region are the background $s$-wave scattering length $a_\mathrm{bg}$ for the open entrance channel, the resonance width $\Delta$, and the difference $\delta \mu$ in magnetic moments between the separated atoms in the entrance channel and the magnetic moment of the ''bare'' resonance state in the closed channel (the ''bare'' state is the approximate eigenstate in the closed channel before coupling to the entrance channel is turned on).   Julienne and Gao~\cite{Julienne2006} have shown how the Mies version of MQDT~\cite{Mies1984a,Mies1984b} can be can be adapted for the Feshbach resonance scattering states.  MQDT theory first develops a set of uncoupled ''reference'' states with which to analyze the problem.  Each 'reference' channel $\beta$ for a given partial wave is characterized by a single reference potential that dissociates to energy $E_\beta$ as $R \to \infty$ and which has a reference scattering phase shift $\eta_\beta(E)$ for $E \ge E_\beta$ and a reference bound state phase shift $\nu_\beta(E)$ for $E<E_\beta$.  Bound states exist in the reference channel for a discrete set of energies for which $\tan\nu_\beta(E) =0$.  Near threshold, two auxiliary MQDT functions $C_\beta(E)$ and $\tan\lambda_\beta(E)$ are needed for each reference channel to characterize the quantum threshold behavior.  These auxiliary functions can be physically interpreted using semiclassical concepts~\cite{Mies1984a,Mies1984b,Julienne1989} and have the property that $C_\beta(E) \to 1$ and $\tan\lambda_\beta(E) \to 0$ when collision energy is sufficiently large.  For $s$-waves in a van der Waals reference potential, this means $E-E_\beta \gg \bar{E}$; semicalssical connections between the short and long range parts of the wave function break down when $E-E_\beta < \bar{E}$, and the quantum connections expressed by the $C_\beta(E)$ and $\tan\lambda_\beta(E)$ functions need to be used.

Even in a complex multichannel problem, it is usually sufficient to represent the closed channel bound state as a single bound state whose properties do not change with magnetic field detuning over a modest range of detuning~\cite{Kohler2006,Mies2000,Nygaard2006}.  Thus we can reduce the tunable Feshbach resonance problem to an effective 2-channel problem with a single ''bare'', or uncoupled, closed channel bound state $|c\rangle$ and a single entrance channel we call the ''background'' channel, $\alpha=bg$.  With this framework and using the background channel van der Waals potential as the reference channel, the scattering phase shift for the fully coupled problem, where the ''bare'' closed channel with bound state energy $E_c=\delta \mu(B-B_c)$ interacts with the entrance channel with background phase shift $\eta_\mathrm{bg}(E)$ to make a scattering resonance, the phase shift $\eta(E,B)$ for $E>0$ is~\cite{Julienne2006}
\begin{equation}
   \eta(E,B)=\eta_\mathrm{bg}(E)-\tan^{-1}\left(\frac{\frac{1}{2} \bar{\Gamma}_c C_\mathrm{bg}(E)^{-2}}{E -\delta \mu(B-B_c)-\frac{1}{2}  \bar{\Gamma}_c \tan\lambda_\mathrm{bg}(E)}\right ) \,,
   \label{eq:MQDTeta}
\end{equation}
where the coupling between the closed and background reference channels is completely contained in the coupling parameter $ \bar{\Gamma}_c $, which is independent of energy $E$ and magnetic field $B$ as these vary near the resonance over ranges on the order of $\bar{E}$ and $\Delta$ respectively.  The resonant phase shift is characterized by an energy-dependent width $\Gamma_c(E)=\bar{\Gamma} C_\mathrm{bg}(E)^{-2}$ in the numerator and an energy-dependent shift $\delta E_c(E)=\delta \mu \delta B_c(E) = (\bar{\Gamma_c}/2)\tan{\lambda_\mathrm{bg}(E)}$ in the denominator of the resonance term in Eq.~(\ref{eq:MQDTeta}).     The general form of Eq.~(\ref{eq:MQDTeta}) has been shown to be in excellent agreement with near-threshold coupled scattering calculations for a number of examples of Feshbach resonances in the literature~\cite{Julienne2006}.   Close to threshold the MQDT functions have the following $s$-wave limiting forms, which permit analytic limiting expressions to be given for Feshbach scattering as $E \to 0$~\cite{Mies2000b}: $C_\mathrm{bg}(E)^{-2} \to k\bar{a}  (1 +(r-1)^2$, $ \tan\lambda_\mathrm{bg}(E) \to 1-r$, where $r=a_\mathrm{bg}/\bar{a}$ is the scattering length in dimensionless $\bar{a}$ units.

In a similar way, the MQDT coupled channels bound state equation from Refs.~\cite{Mies1984a,Mies1984b} can be put in the form
\begin{equation}
\label{eq:MQDTbound}
 \left (  E-\delta \mu (B-B_c) \right )  \tan \nu_\mathrm{bg}(E) = \frac{\bar{\Gamma}}{2} \,.
\end{equation}
where $\nu_\mathrm{bg}(E)$ is the background reference channel MQDT phase function for $E<0$, which has the property near threshold~\cite{Mies2000b} that $ \nu_\mathrm{bg}(E) \to \tan^{-1}[1/(r-1)] -\bar{a}\kappa$, where $E=-\hbar^2 \kappa^2/(2\mu)$ for $E<0$.   Using this threshold property, it is straightforward to show that the coupled channels bound state crosses threshold at $B_0=B_c+\delta B_c$, where the shift $\delta B_c = -\Delta \, r(r-1)/(1+(r-1)^2)$ is the same shift that Eq.~(\ref{eq:MQDTeta}) gives for the singularity at $B=B_0$ in the scattering length $a(B)$; see Eq.~(\ref{eq:aB}).

One advantage of the MQDT approach is the factorization of the resonance coupling into a part associated with the long-range physics, $C(E)^{-2}$, and a reduced coupling parameter, $\bar{\Gamma}$, associated with the short range physics.    The $\bar{\Gamma}$ factor is proportional to the dimensionless MQDT energy-insensitive $Y_\mathrm{c,bg}$ matrix element which gives the strength of the coupling:
\begin{equation}
  |Y_\mathrm{c,bg}|^2 = \frac{\bar{\Gamma}}{2} \frac{\partial \nu_c(E)}{\partial E} \,,
\end{equation}
where ${\partial \nu_c(E)}/{\partial E}\approx \pi/\Delta E_c$ and $\Delta E_c$ is the mean vibrational spacing between adjacent vibrational levels near $E=E_c$.  For all ultracold atom Feshbach resonances in the literature, $Y_\mathrm{c,bg}\ll1$ represents weak coupling, that is, the width of the resonance is small compared to the spacing between adjacent bound vibrational levels in the closed channel (mixing sometimes occurs with other closed channels; see Figure  8).

 In order to more explicitly show the connection between the long-range potential and Feshbach resonance properties, Chin {\it et al.}~\cite{Chin2008} introduced a dimensionless resonant strength parameter $s_\mathrm{res}$ by expressing $a_\mathrm{bg}$ and $\delta \mu \Delta$ in dimensionless units of $\bar{a}$ and $\bar{E}$ respectively:
\begin{equation}
s_\mathrm{res} = \frac{a_\mathrm{bg}}{\bar{a}} \frac{\delta \mu \Delta}{\bar{E}} \,.
\label{eq:sres}
\end{equation}
This is the inverse of the $\eta$ parameter used by Ref.~\cite{Kohler2006} to characterize Feshbach resonances.  The resonance width $\Gamma_c(E)=(k\bar{a})(\bar{E} s_\mathrm{res})$ exhibits the threshold energy-dependence near $E=0$, and the energy-insensitive MQDT resonance strength $\bar{\Gamma}$ is
\begin{equation}
 \frac{\bar{\Gamma}}{2} = \bar{E} \frac{s_\mathrm{res}}{1+(r-1)^2} \,.
 \label{eq:Gammabar}
\end{equation}
This expression can be used in Eq.~(\ref{eq:MQDTbound}) to get a general dimensionless MQDT equation for bound states.  The bound state equation can be used to show that as $\kappa \to 0$ the bound state energy is given by $\kappa = 1/(a(B)-\bar{a})$, which is the same relation for large positive scattering length as found by Gribakin and Flambaum~\cite{Gribakin1993} for a single van der Waals potential, namely, $\kappa_\mathrm{bg} = 1/(a_\mathrm{bg}-\bar{a})$ when $a_\mathrm{bg} \gg \bar{a}$.  In the Feshbach resonance case, the relation applies when $a(B) \gg \bar{a}$ when $s_\mathrm{res} \gg 1$, but only applies over a much more restricted range where $a(B) \gg 4\bar{a}/s_\mathrm{res} $ when $s_\mathrm{res} \ll 1$.  Solving Eq.~(\ref{eq:MQDTbound}) in general gives the right coupled channels bound states; see Ref.~\cite{Chin2008}.  When $B-B_0$ is sufficiently far from resonance, the equation also recovers the ''bare'' bound state energy varying as $E = \delta \mu (B-B_c)$.  The variation with $B$ of the bound state energy of the coupled Feshbach bound state can be used to determine the norm $Z$ of the closed channel piece in the unit normalized coupled channels bound state wave function: $Z=\delta \mu^{-1} \partial E/ \partial B$~\cite{Kohler2006,Chin2008}.

The dimensionless parameter $s_\mathrm{res}$ permits the classification of resonances into two basic types~\cite{Kohler2006,Chin2008}, depending on how their character changes as $B$ is tuned over a range on the order of $\Delta$ near $B=B_0$.  One type with $s_\mathrm{res} \gg 1$ is ''open channel dominated'' and the other type with $s_\mathrm{res} \ll 1$ is ''closed channel dominated''.  In either case this pertains to the character of the bound state over a tuning range that is a significant fraction of $\Delta$.  In the open channel dominated case,  the wave function takes on the spin character of the entrance channel,  has a universal energy given by Eq.~(\ref{eq:aB}), and closed channel norm $Z \ll 1$.  In contrast, the closed channel dominated case has only a very small domain of universality very close to $B_0$ and takes on the spin character of the closed channel bound state with $Z \approx 1$.  Some of the more interesting experimental resonances are open channel dominated, namely the $^6$Li 83.4 mT, $^{40}$K 20.2 mT, and $^{85}$Rb 15.5 mT resonances, but many others are closed channel dominated~\cite{Kohler2006,Chin2008}.

The $^{40}$K$^{87}$Rb 54.69 mT resonance, for which $\delta \mu = 2.40\mu_\mathrm{B}$ and $s_\mathrm{res}=2.07$, is an example of a resonance tending to open channel dominance.  Thus, Figure 8 shows that it has a universal domain over about a third of its width, and in fact, a calculation of $Z$ shows that $Z < 0.4$ for detuning up to 0.1 mT.  At 54.6 mT near the field of the experiment of Ni {\it et al.}~\cite{Ni2008} the energy is $E/h=-329$ kHz, $Z=0.28$, and the scattering length is 443 a$_0$, predicting a ''universal'' energy of -336 kHz.  Figure 10 shows that the open channel component $f_{aa,aa}(R)$ of the coupled channel wave function at 54.6 mT extends to very large $R \gg \bar{a}$.  This entrance channel component has a norm of 0.718 and is the dominant part of the wave function.  The wave function becomes even more ''open channel dominated'' as $B$ increases towards resonance at $B_0$.  It will show more closed channel character as $B$ tunes farther away from resonance.

Figure 10 shows the $f_{rb,aa}(R)$ closed channel component of the bound state wave function at 54.2 mT, where the energy of the bound state is $E/h = -10.6$ MHz, the detuning from resonance is about 1.6 resonance widths $|\Delta|$, and the scattering length is $-71$ a$_0$.  This level is far from resonance, the bound state is far from universality, and the norm of the $f_{rb,aa}(R)$ closed channel component is 0.90.  It is evident that the $f_{rb,aa}(R)$ component has a very strong overlap with the $n=-2$ bound vibrational level of the long range potential, so that it is legitimate to characterize the level to a good approximation as being an $n=-2$ bound state of the $rb$ channel.  The other $rf(-3d0)$ resonance in Figure 8 is a closed channel dominated resonance with $s_\mathrm{res}=0.12$.  It is a very narrow resonance with a very small ($\ll \mu$T) domain of universality, so the below-threshold bound state is very strongly of $d$-wave $n=-3$ $rf$ channel character. 
 
Using the full MQDT in its angular momentum insensitive form~\cite{Gao2005} should permit the prediction of the positions and widths of the various Feshbach resonances in all the channels of the problem.   The only additional information needed to develop this theory is the analytic basis set transformation (''frame transformation'') between the short-range and long-range basis sets for representing the approximate spin eigenstates of the system.  This method is similar to the asymptotic bound state (ABM) method used by Wille {\it et al.}~\cite{Wille2007} to characterize $^6$Li$^{40}$K resonances.  The MQDT method is an ''on-the-energy-shell" coupled channels method, whereas the ABM method relies on an expansion in a basis set of bound states of the X$^1\Sigma^+$ and a$^3\Sigma^+$ potentials.  Either are capable of giving an approximate coupled channels representation of the bound states of the system, and finding the threshold resonances within the framework of those approximations.  Either method could provide a useful alternative to full numerical coupled channels calculations.

\subsection{Inelastic collision rates}

Finally, there is an additional insight from MQDT that is very helpful in estimating, and in some cases quantitatively calculating, the rate coefficients for ultracold atomic or molecular collisions.  The basic idea of the method is given in Ref.~\cite{Julienne1989} and was implemented for atomic collisions with Penning ionization by Refs.~\cite{Orzel1999,Dickinson2007} and for molecular vibrational quenching collisions by Ref.~\cite{Hudson2008}.  It takes advantage of Eqs.~(\ref{eq:MiesK})-(\ref{eq:MiesfD}) together with the MQDT concept of factoring the $S$-matrix into separate parts due to short- and long-range interactions, as we saw with the resonant width in the last section.  

Let us assume that the probability for loss of the colliding species A and B is unity, given that the species are close together and strongly interacting at distances small compared to the scale length $\bar{a}$ of the long range potential.  If the collisional wave function were semiclassical at all $R$ and not affected by quantum threshold effects, then one could calculate a semiclassical rate coefficient in a Langevin model, summing over all contributing partial waves that reach short distance, taking $1-|S_{\alpha\alpha}|^2 =1$  in Eq.~(\ref{eq:MiesfD}) to represent maximal loss and no reflection for the contributing partial waves.  However, in the threshold regime, the threshold law associated with the long De Broglie wavelength needs to be taken into account.  Let us assume that collision energy $E$ is low enough and no electric field is present to induce an actual molecular dipole moment so that only $s$-wave contribute.  The total $s$-wave loss is calculated by taking the loss probability $1-|S_{\alpha\alpha}|^2 $ to be the transmission probability for reaching short range, namely, $1-P_r$, where $P_r$ is the quantum reflection probability from the long-range potential.  This gives the needed quantum correction to semiclassical theory as $E \to 0$.  The transmission factor can be calculated numerically~\cite{Orzel1999,Hudson2008} or, in some cases, analytically~\cite{Friedrich2004,Dickinson2007,Cote1997}.  Given the analytic result $1-P_r=4k\bar{a}$ for $s$-wave transmission through the long range van der Waals potential~\cite{Friedrich2004,Cote1997},  the rate coefficient for collisional loss from Eq.~(\ref{eq:Kin}) with $b=\bar{a}$ is 
\begin{equation}
 K_\mathrm{in}=2(h/\mu)\bar{a}
 \label{eq:Kuniversal}
 \end{equation}
 for a ''strong'' molecular loss collision with unit short range probability of an inelastic event.   This formula is an example of ''inelastic universality'', where there is a universal rate constant depending only on the long-range potential and not on the scattering length, since there is no ''back reflection'' from the short-range region.  Typical values of $\mu$ and $\bar{a}$ give an order of magnitude of $10^{-10}$ cm$^3/$s for $K_\mathrm{in}$ for such collisions. 

Consequently, if the probability for a short-range loss process is very high, then knowing the long range van der Waals coefficient is sufficient to make quite accurate estimates of the threshold collision rate coefficient, as demonstrated by applying the above model to vibrational quenching of excited vibrational levels of the RbCs molecule~\cite{Hudson2008}.  This factorization procedure of MQDT theory~\cite{Julienne1989} can be generalized to include other partial waves or resonance structure~\cite{Machholm2001} and should be capable of giving much insight into molecular collisions.  If the probability $\bar{P}$ of short-range reaction or loss is near unity and one has a way to calculate or estimate it, then that factor can be included in the expression for the $s$-wave rate constant by replacing $\bar{a}$ in Eq.~(\ref{eq:Kuniversal}) by $\bar{a}\bar{P}$.  On the other hand, if $\bar{P} \ll 1$ so there is significant reflection from the short range region, $K_\mathrm{in}$ will depend on the scattering length $a_\alpha$ of the entrance channel.    In any case, we see that if a collision leads to a ÓstrongÓ highly probable short range loss event, the influence of the long-range potential is critical in determining how threshold modifications to Langevin theory occurs in the ultracold domain.

\section{Deeply bound states of $^{40}$K$^{87}$Rb}\label{deep}

So far we have concentrated on bound states that are very close to threshold with binding energies of less than 1 GHz.  Figures 11 - 13 show an expanded view of the $M=-7/2$ $s$-wave bound states with binding energies up to 30 GHz, or about 1 cm$^{-1}$, between $B=$ 0 and 120 mT.  The character of the vibrational levels is relatively easy to explain in this region of the spectrum, where the splitting between adjacent vibrational levels of $^1\Sigma^+$ and $^3\Sigma^+$ symmetry is much smaller than the splitting of the atomic hyperfine manifold.  The singlet and triplet vibrational bound states have similar binding energies because they (accidentally) have similar scattering lengths.  Each vibrational level, away from avoided crossings where different levels mix, is characterized by the dominant spin character of one of the 12 separated atom channels $\beta$ shown in Figure 6.  They also are characterized by a vibrational wave function characteristic of $n=-1,-2,\dots$ long range levels, so that levels are located at energies that are below the corresponding separated atom limits by the binding energies of the $n$ levels in these channels.  Figure 11 indicates these groupings, and shows how the levels in the close-up views in Figures 6 and 7 relate to neighboring levels.  Levels of different channels and $n$ are intermixed because of the similarity in vibrational spacings and hyperfine spacings in some cases.

Figure 12 extends the broader view to a scale of 10 GHz.  The A and B groups cluster in the 10 GHz region, since the binding energy of the $n=-3$ levels are on the order of 10 GHz.  The $aa(-3)$ level parallel to the $E=0$ axis has the spin character of the $aa$ entrance channel and is the level made by Ospelkaus {\it et al.}~\cite{Ospelkaus2008} by 2-color STIRAP transfer of population from the Feshbach molecule state at 54.6 mT.

Figure 13 gives an even more expanded view down to 30 GHz binding energy, which includes the full manifold of the 12 $n=-4$ levels associated with each of the separated atom channels  $\beta$.  The Figure also shows the energy levels of the X$^1\Sigma^+$ and a$^3\Sigma^+$ potentials, referenced to the nonrelativistic energy of the separated atoms at $B=0$.  This energy is 4.843439 GHz above the energy of two $a$ atoms in the $aa$ channel at $B=0$.    In this domain of binding energy, the Figure shows that the splitting between the two adjacent singlet and triplet levels is much less than the splittings of the atomic hyperfine manifold.  Each molecular level, except near avoided crossings of levels of different channels, thus has the $\beta n$ character of two weakly bound separated atoms in channel $\beta$ with the long-range vibrational wave function of level $n$.  If the singlet and triplet potentials had very different scattering lengths, so the corresponding singlet and triplet vibrational levels were not so close in energy as in the Figure, then one would begin to see a breakdown in this atomic coupling scheme in this range of energy.

Figures 14 and 15 examine the transition from the threshold domain, where the singlet and triplet levels are mixed by spin-dependent interactions, and the more deeply bound region, where the vibrational levels are to a good approximation dominantly singlet or triplet in character.  This transition occurs for binding energies of around 200 GHz, where the splitting between adjacent $n=-8$ singlet and triplet vibrational levels at $-205.987$ GHz and $-192.174$ GHz becomes on the order of the atomic hyperfine splittings.  Levels with less binding tend to be of mixed singlet and triplet character.  More deeply bound levels are clearly identified as being dominantly singlet or triplet, with the exception that the $n=-13$ singlet levels at $-922.325$ GHz has an accidental near-degeneracy with the $n=-14$ triplet level at $-924.070$ GHz.  This gives rise to a strong mixing of the two levels that is very local in energy.  Singlet-triplet mixing is very small for all levels with larger binding energy.

Finally, Figure 16 shows the spin structure of the $v=0$ $J=0$  X$^1\Sigma^+$  ground state molecule.    Since the two active electrons are paired into a singlet state with no net spin, and since the projection on the molecular axis $\Lambda=0$ for a $\Sigma$ molecular state, there is no coupling of the nuclear spins to the electrons or the molecular axis.  If there were no coupling to distant triplet electronic states, all nuclear spin components of the $v=0$ $J=0$  X$^1\Sigma^+$ level would be degenerate.  Figure 16 shows a slight zero-field splitting at $B=0$ of 41 kHz 
between the lowest energy $I_\mathrm{tot}=11/2$ and highest energy $I_\mathrm{tot}=5/2$ nuclear spin components, where $I_\mathrm{tot}$ is the resultant of the $^{40}$K and $^{87}$Rb nuclear spins.  This very small splitting in our model is due to second-order coupling through the distant a$^3\Sigma^+$ state.  This model calculation based on atomic spin coupling constants may not be accurate in the molecular environment of the spins in the $v=0$ level.  However, the nuclear spin structure at the 54.6 mT field of the experiment of Ref.~\cite{Ni2008} represents essentially uncoupled nuclear spins in the large $B$ field, where the energies are very close to being the energies of the isolated separate atoms in the same field.  Thus, the $^{40}$K atom splits into 9 Zeeman components and the $^{87}$Rb atoms splits into 4 components with a total spread of 3.4 MHz across the manifold of levels at 54.6 mT.

It is a very interesting question to determine to what extent individual nuclear spin states can be prepared and manipulated using the very precise optical control that may be possible with the ultra high precision frequency comb technology that went into the STIRAP experiment of Ni {\it et al.}~\cite{Ni2008}.  This may be possible using a combination of polarization and frequency control.  There also is the question of the collisional stability of the $v=0$ ground state molecules.  If reactive collisions are inhibited by a reaction barrier at $\mu$K temperatures, then the only destructive collisions that change the state are ones that depolarize or relax the nuclear spins.  Even at zero field, the spread of energy of 41 kHz corresponds to $E/k_B$ of 2 $\mu$K.  Due to the absence of coupling of the nuclear spins to the electrons or molecular axis in the problem, spin depolarization due to molecule-molecule collisions probably has a very small inelastic loss rate constant (perhaps too small to be measurable).  However, a collision with a $^{40}$K or $^{87}$Rb atom could have a non-negligible inelastic collision rate coefficient, due to the coupling of the nuclear spins in the molecule to the unpaired atomic electron during the collision.  It will be important to determine if the nuclear spins can be controlled in a polar molecule gas or lattice.  For example, the spins might make good quantum memory, subject to weak decoherence, but capable of rapid optical manipulation and measurement.

It is worth noting the excellent quality of the potentials in Ref.~\cite{Pashov2007} that were used in this calculation.  We calculate the energy of the $v=0$ $J=0$  X$^1\Sigma^+$ level relative to the energy of two $a$ state atoms at 54.588 mT  to be $E/h = -125320.10$ GHz.  This differs by only $0.4$ GHz from the measured value of $-125319.703(1)$ GHz, which corresponds to an error of only 3 parts in $10^6$ in the absolute binding energy.  The calculations for the a$^3\Sigma^+$ state are not as accurate.  The lowest spin component of the $v=0$ $N=0$ triplet state was measured~\cite{Pashov2007} to be at  $-7180.4180(5)$ GHz, compared to a calculated value of $-7195.6$ GHz.  The much larger error of 16 GHz is likely due to the lack of accurate spectroscopic data for the lowest vibrational levels to constrain the minimum of the a$^3\Sigma^+$ potential.  The new very precise data on the absolute molecular binding energies should aid the construction of much more accurate potentials for future use.

\section{Concluding remarks}

This paper has presented some very general considerations for ultracold atomic and molecular collisions and interactions as well as some specific results for the $^{40}$K$^{87}$Rb fermionic molecule.  A general consideration is that ultracold collisions very precisely prepare the collision complex of two colliding species in a very sharp energy state, which then serves as a gateway for various kinds of spectroscopic probing and manipulation of the complex.  Thus, one can manipulate the properties of an ultracold gas or lattice using tunable scattering resonances, one can use such resonances to make a near-threshold bound state, and one can probe or populate levels far from threshold by optical manipulation.  Thus, resonant scattering theory and an account of near-threshold bound states are of crucial importance to ultracold atomic and molecular physics.  

Much progress can be made by understanding  the spectrum and resonances associated with the long-range potential.  Good theoretical progress in understanding the more complex world of molecular collisions can be made by developing theories that explicate the near-threshold domain of resonant scattering.  Much of the physics associated with strong short-range interactions can be parameterized by energy-insensitive parameters of multichannel quantum defect theory.   A simple example of applying these ideas is in understanding the magnitude of rate constants for inelastic molecular vibrational relaxation, where in the ultracold domain, one only needs to know the quantum reflection off the long-range potential between the two colliding species, given that relaxation has unit probability if the colliding species reach the short-range region.  Such ideas can be generalized to include resonances and weak short range processes.  It remains to be seen how successful such an approach can be in the more complex environment of molecular collisions.

The bound and threshold scattering states of the $^{40}$K$^{87}$Rb molecule illustrate some important aspects of molecular physics related to the formation and use of a vibrational ground state molecule.  The domain from near-threshold to the most deeply bound levels can be understood quite precisely from coupled channels calculations.  These require the full Hamiltonian for the problem, including accurate molecular adiabatic Born-Oppenheimer potentials and spin coupling constants.  In general these can only be obtained from a combination of accurate {\it ab initio} calculations and precise spectroscopy.  While excellent quality potentials are available for KRb, these will need to be developed of other systems of future interest.  Much insight and practical calculation can be done in the near-threshold domain from a combination of methods involving the properties of the long-range potential and various implementations and approximations based on the long-range potential.  These methods require a semi-empirical approach where parameters of the theoretical models, such as scattering lengths for the potentials, need to be extracted from fitting experimental data.  Such methods are already highly developed for atomic collisions, but still need to be developed for molecular collisions, if possible.   It is clear, however, that accurate theoretical models of the spin structure of ground state molecules will be needed, including the effect of external magnetic, electric, or electromagnetic fields.  Ultracold resonant scattering theory of colliding molecular species with structure must first understand the various scattering channels associated with the internal structure of the colliding species, and then try to understand the resonances or other scattering properties associated with the long-range potential.  Finally, the threshold resonances and dynamics associated with strong short-range interactions of the molecular collision complex will be difficult to predict accurately from first principles, but may be accessible to a combination of theoretical modeling and experimental probing of the complex.

%The references should start on their own page.

\clearpage

\bibliographystyle{fd-bibtex}
\bibliography{Faraday}

%Please compile a list of all figure captions on a separate page:
%Figure captions go here

\clearpage

\begin{figure}[ht]
  \begin{center}
  \includegraphics[angle=0,width=12cm]{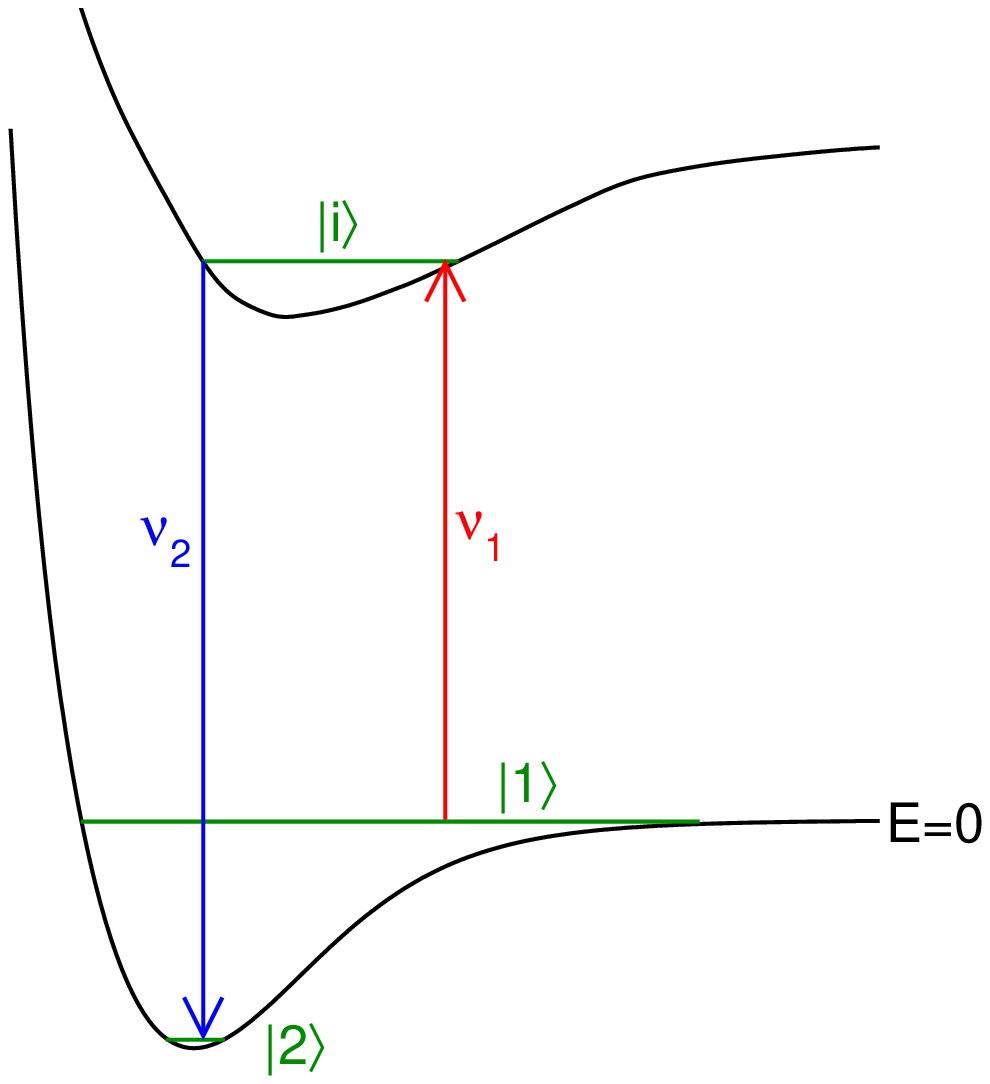}
   \caption{Schematic figure of the potential energy curves $V(R)$ of two interacting atoms A and B versus interatomic distance $R$, with the zero of energy $E$ set as the energy of two separated  atoms in the states in which they are prepared.  The horizontal line just below threshold indicates the energy of a weakly bound molecular level $|1\rangle$, which is prepared by associating two cold atoms and then converted to the target deeply bound vibrational level $|2\rangle$ by a Raman transition through an excited intermediate molecular state $|i\rangle$.  The Raman process could also be initiated starting from a scattering state with $E >0$, although this has proved difficult in practice.  Schemes of this general type were proposed by References~\cite{Kokkelmans2001,Jaksch2002,Damski2003} and realized in References~\cite{Sage2005,Danzl2008,Ni2008} by starting from associated atoms in a weakly bound state.}
  \end{center}\label{Fig1}
\end{figure}

\clearpage

\begin{figure}[ht]
  \begin{center}
   \includegraphics[angle=270,width=\columnwidth]{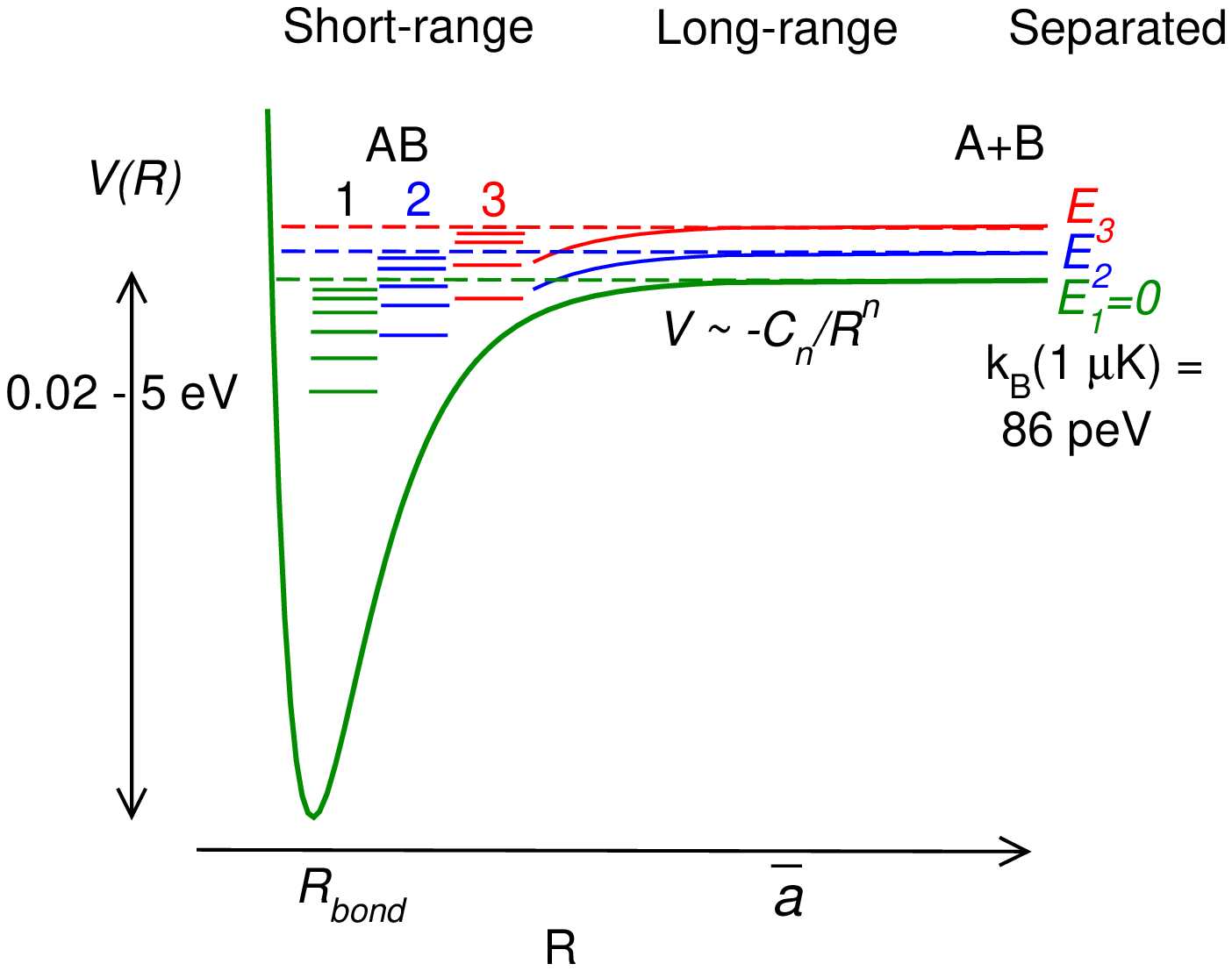}
   \caption{Schematic view of an ultracold collision to form a molecular "complex".  The species A and B are each prepared in internal states labeled collectively by $\alpha=1$ and with a very small relative collision energy near the separated species energy $E_1=0$.  Two closed channels 2 and 3 with different internal energies $E_2$ and $E_3$ of the separated species are also schematically indicated, along with the spectrum of associated bound state levels (short horizontal lines) in channels 1, 2 and 3.  The energy scale for an ultracold collision is on the order of $k_BT=86$ peV or $k_B T/h = 21$ kHz for $T=1$ $\mu$K.  The energy scale for the short range part of the potential, where $R \approx R_\mathrm{bond}$ in on the order of the chemical bond length $R_\mathrm{bond}$, is given by the binding energy of the $v=0$ level of the ground state potential, on the order of $E/h=$5 THz for a weakly bound van der Waals molecule (0.02 eV) to 1 PHz for a strong chemical bond (5 eV).   The near-threshold bound and scattering states of the complex are sensitive to the long range part of the potential that varies asymptotically as $-C_n/R^n$ and has a characteristic length $\bar{a}$.  Near-threshold bound states associated with closed channels 2 and 3 of the collision form threshold scattering resonances in channel 1 that can be tuned across $E=E_1$ by varying magnetic, electric, or electromagnetic fields.}
  \end{center}\label{Fig2}
\end{figure}

\clearpage

\begin{figure}[ht]
\label{Fig3}
  \begin{center}
   \includegraphics[angle=270,width=\columnwidth]{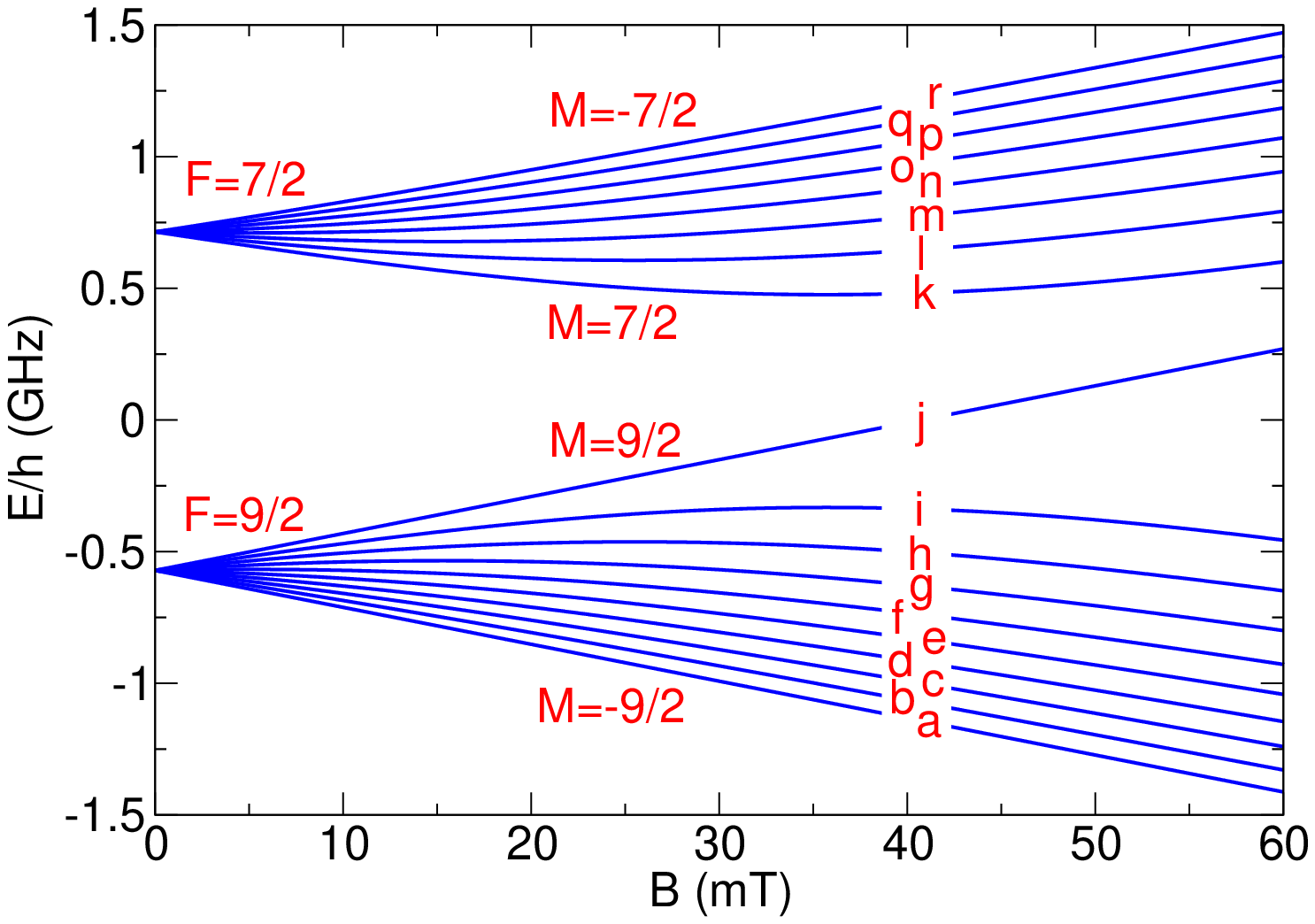}
   \caption{The energy of the magnetic Zeeman sublevels of the fermionic $^{40}$K atom versus magnetic field $B$.  The zero of energy is the nonrelativistic energy center of gravity of the multiplet.  The levels, labeled $a,b,\ldots$ in order of increasing energy correlate at $B=0$ with the two hyperfine levels with total electron spin plus nuclear spin angular momentum $F=9/2$ and $F=7/2$. The $M$ quantum number specifies the angular momentum projection on the magnetic field axis and, unlike $F$, remains a good quantum number as $B$ increases.}
  \end{center}
\end{figure}

\clearpage

\begin{figure}[ht]
  \begin{center}
   \includegraphics[angle=270,width=\columnwidth]{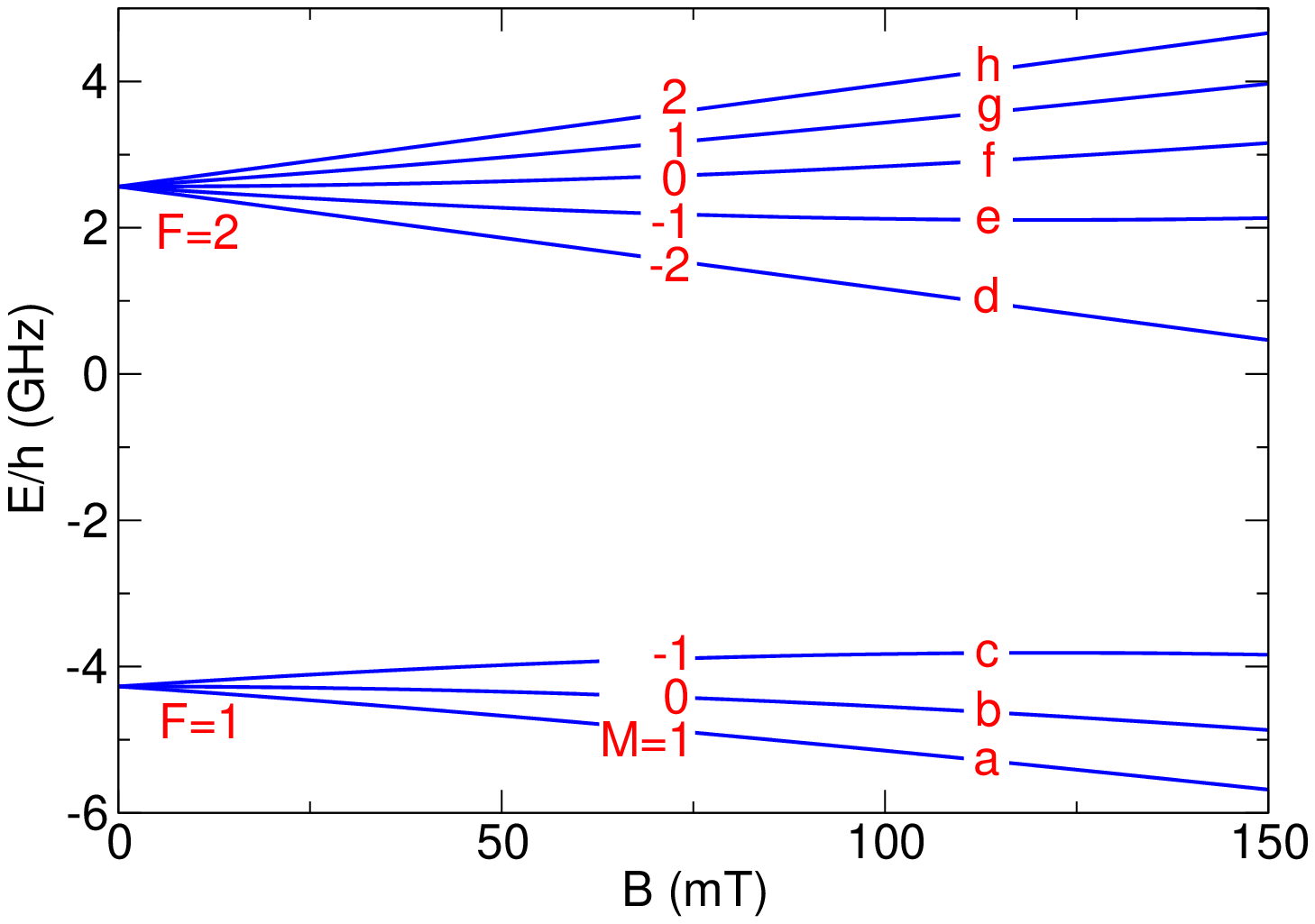}
   \caption{The energy of the magnetic Zeeman sublevels of the bosonic $^{87}$Rb atoms versus magnetic field $B$.  The zero of energy is the nonrelativistic energy center of gravity of the multiplet. The levels, labeled $a,b,\ldots$ in order of increasing energy correlate at $B=0$ with the two hyperfine levels with total electron spin plus nuclear spin angular momentum $F=1$ and $F=2$. The $M$ quantum number specifies the angular momentum projection on the magnetic field axis and, unlike $F$, remains a good quantum number as $B$ increases. }
  \end{center}\label{Fig4}
\end{figure}

\clearpage

\begin{figure}[ht]
  \begin{center}
   \includegraphics[angle=270,width=\columnwidth]{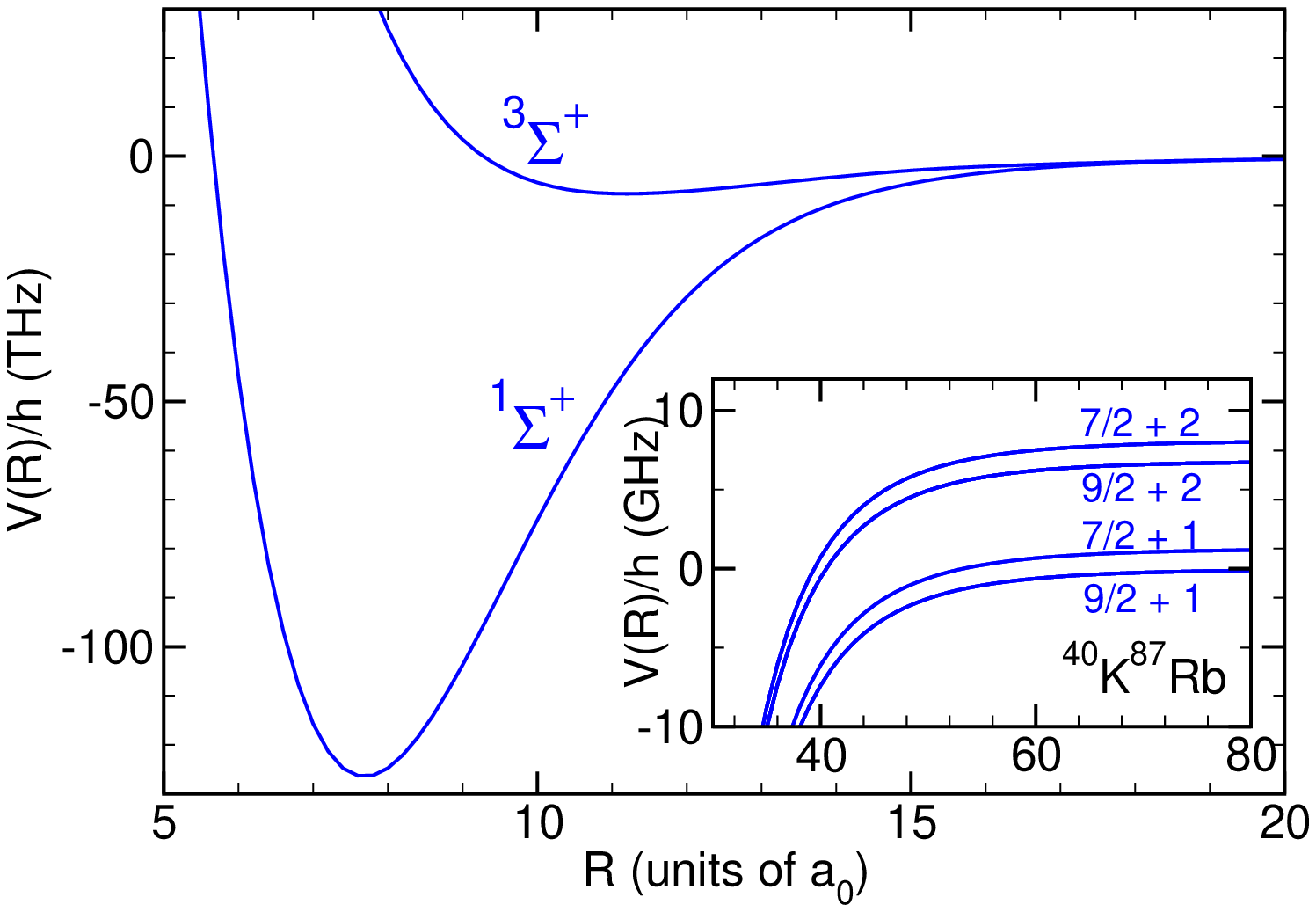}
   \caption{Adiabatic Born-Oppenheimer X$^1\Sigma^+$ and a$^3\Sigma^+$ potential energy curves $V(R)$ of the KRb molecule correlating with $^2$S ground state K and Rb atoms. These curves dissociate to $E=0$ at the nonrelativistic energy center of gravity of the  $B=0$ atomic hyperfine levels.   The inset shows the long range adiabatic curves for the $M=-7/2$ projection states of the $^{40}$K$^{87}$Rb molecule at $B=0$.  These curves separate asymptotically to the atoms in one of their ground hyperfine levels.  In the inset, the energy zero is set as the energy of the lowest $9/2 + 1$ set of levels at B=0.  The characteristic van der Waals length is $\bar{a}=68.9$ a$_0$.}
  \end{center}\label{Fig5}
\end{figure}

\clearpage

\begin{figure}[ht]
  \begin{center}
  \includegraphics[angle=270,width=\columnwidth]{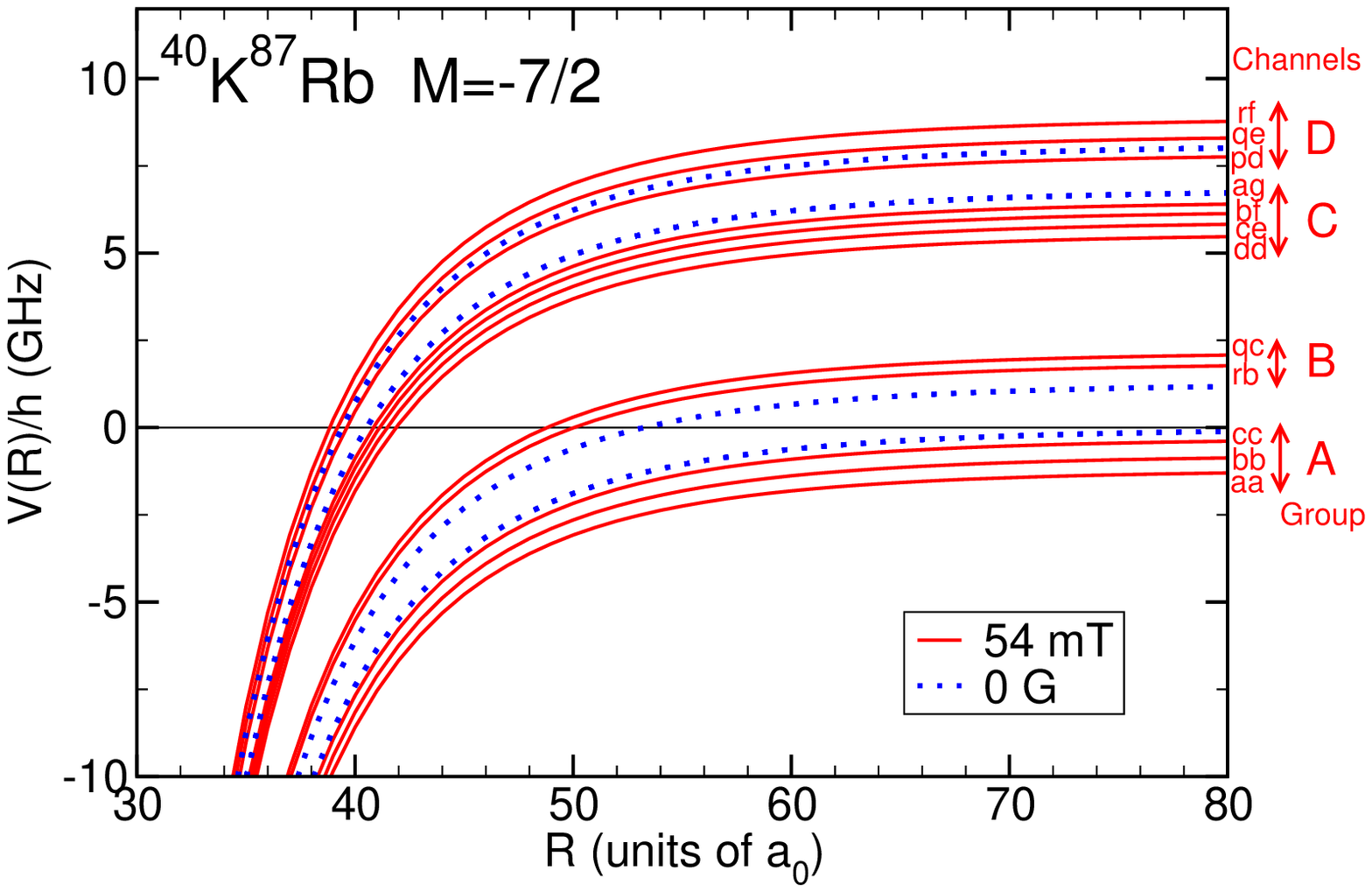}
   \caption{ Long range adiabatic curves for the 12 $M=-7/2$ projection states of the $^{40}$K$^{87}$Rb molecule.  The dashed lines show the same curves as the inset of Fig.~\ref{Fig3} for $B=0$ and the solid lines show the 12 curves for $B=54$ mT, labeled according to the Zeeman levels of each of the separated atoms.  The zero of energy is taken to be the energy of the lowest hyperfine levels $9/2+1$ at zero field.  The groupings into 4 sets of states labeled by $A$, $B$, $C$, and $D$ correspond to the sets of states $(aa,bb,cc)$, $(rb,qc)$, $(dd,ce,bf,ag)$, and $(pd,qe,rf)$.  These respective groupings are associated with the zero field separated atom hyperfine levels $9/2+1$, $7/2+1$, $9/2+2$ and $7/2+2$.  }
  \end{center}\label{Fig6}
\end{figure}

\clearpage

\begin{figure}[ht]
  \begin{center}
  \includegraphics[angle=270,width=\columnwidth]{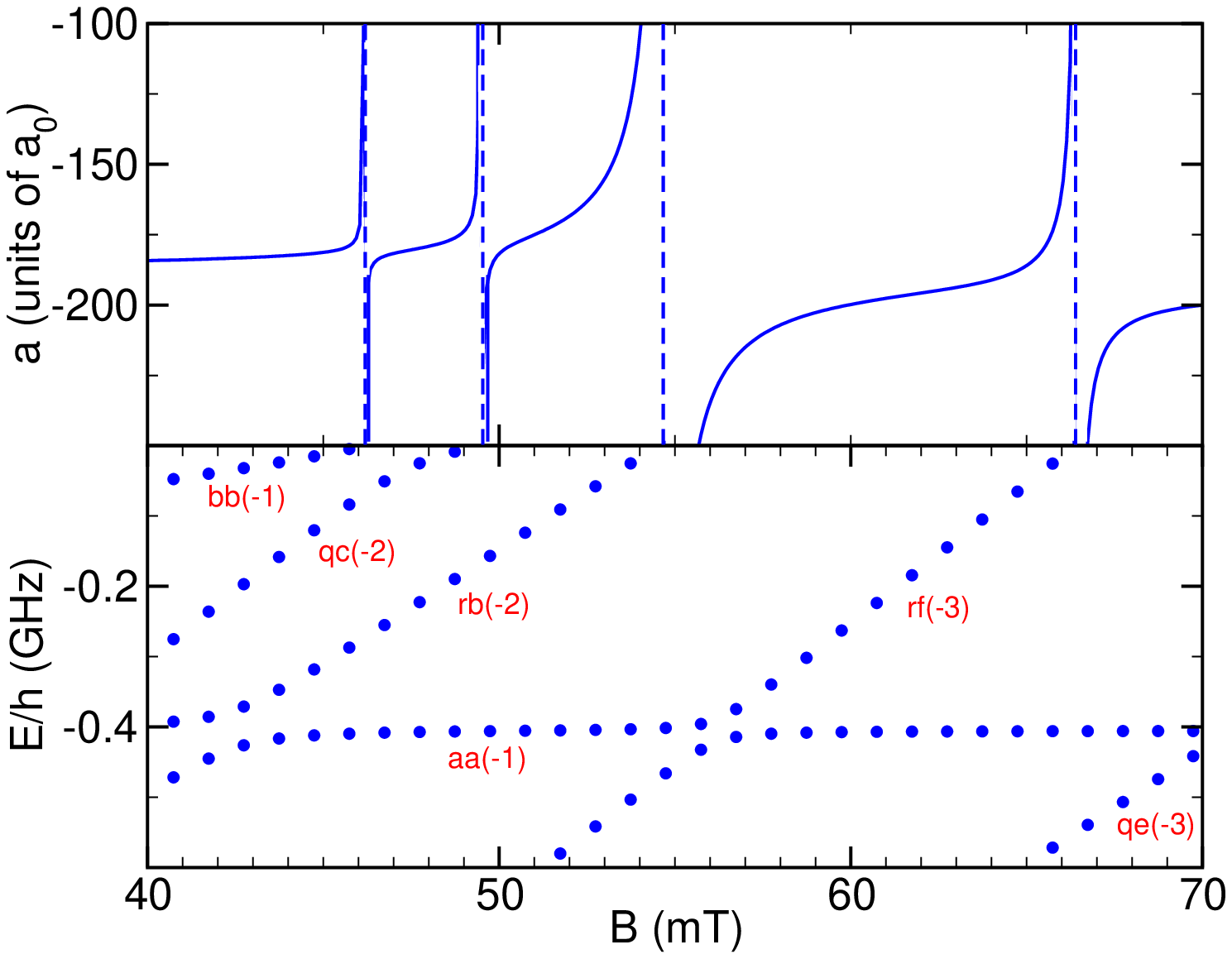}
   \caption{Scattering length (upper panel) and bound state energy (lower panel) for the $M_\mathrm{tot}=-7/2$ $s$-wave channels of the $^{40}$K$^{87}$Rb molecule.  The zero of energy is the energy $E_\alpha$ of the $aa$ channel at each $B$ field.   Thus, the bound state energies give the binding energies of the levels relative to the $aa$ separated atom energy. The dashed vertical lines show the points of singularity $B_0$ of the scattering length, calculated to be at 46.239 mT, 49.563 mT, 54.694 mT and 65.969 mT, where the binding energy of a molecular bound state becomes zero as the state reaches threshold. The solid circles show the bound state energies calculated for a discrete set of $B$ values.   The labels $\beta(n)$ show the dominant spin character of the bound eigenstate, where $\beta$ indicates a separated atom closed channel and $n$ gives the vibrational quantum number of the vibrational level in that channel, counting down from the dissociation limit of the channel at $E_\beta$.}
  \end{center}\label{Fig7}
\end{figure}

\clearpage

\begin{figure}[ht]
  \begin{center}
  \includegraphics[angle=270,width=\columnwidth]{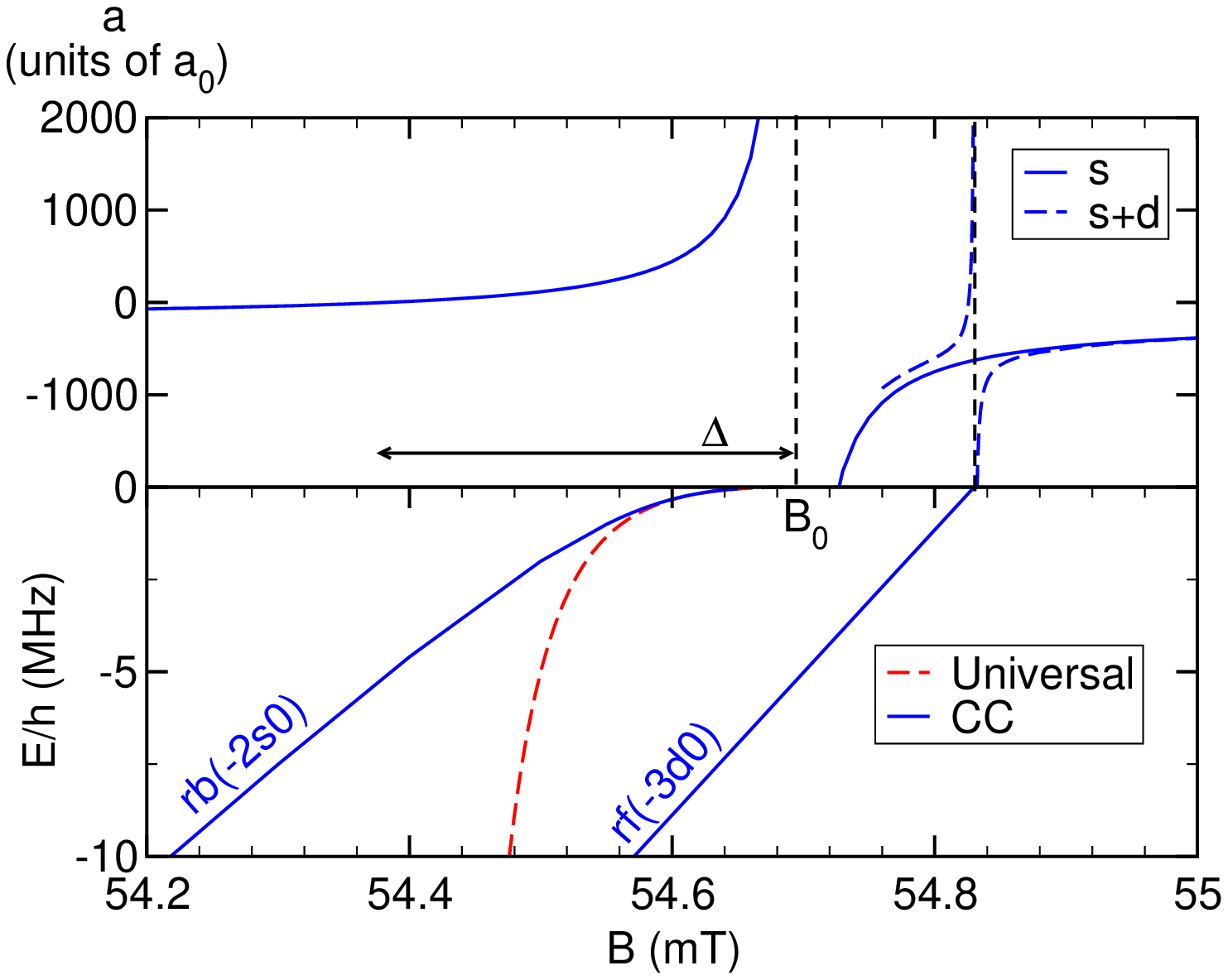}
   \caption{Expanded view of the 54.69 mT resonance in Figure 7.  The upper panel shows the scattering length and the lower panel the bound state energies versus $B$.  The solid line in the upper panel shows the scattering length calculated using an $s$-wave basis set only, the vertical dashed line marks the resonance position $B_0$, and the double headed arrow shows the magnitude of the resonance width $\Delta$.   The dashed curve in the upper panel shows an extra resonance that appears when $s$ and $d$ basis functions are both used in the calculation.  The bound state labeled $rb(-2s0)$ is the same as in Figure 7.  The bound state label includes the $\ell m_\ell = s0$ designation.  The new bound state labeled $rf(-3d0)$ is a $d$-wave bound state that appears when a $d$-wave basis set is added to the calculation with $M_\mathrm{tot}=-7/2$.   The dashed line in the lower panel indicates the universal  energy derived from the scattering length using Eq.~(\ref{eq:aB}).}
  \end{center}\label{Fig8}
\end{figure}

\clearpage

\begin{figure}[ht]
  \begin{center}
  \includegraphics[angle=0,width=\columnwidth]{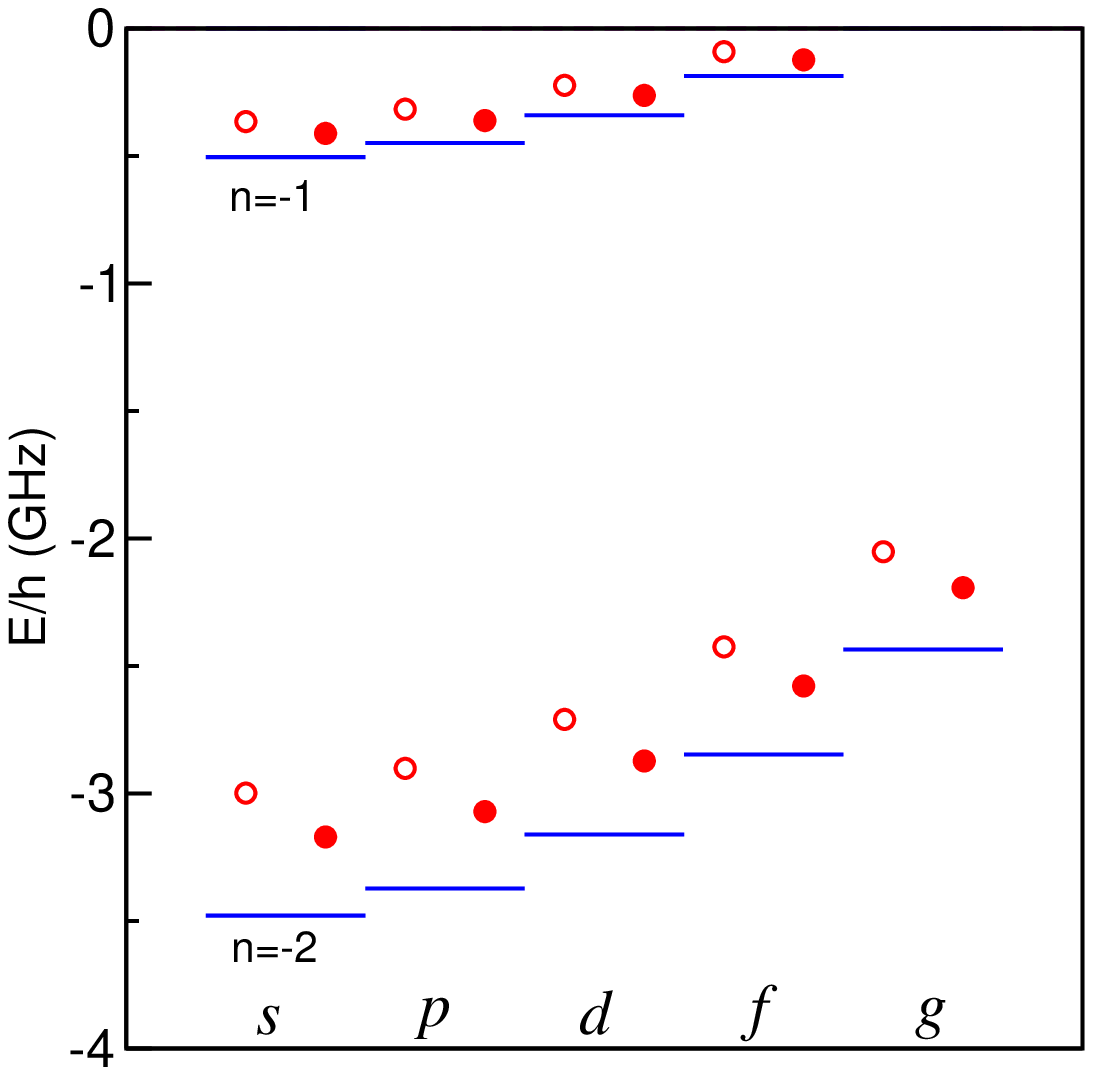}
   \caption{Bound state spectrum of the last two vibrational levels of $^{40}$K$^{87}$Rb with quantum numbers $n=$ $-1$ and $-2$ for partial waves $\ell=0,1,2,3,4$ ($s$,$p$,$d$,$f$,$g$).  The horizontal lines show the levels for a pure van der Waals potential with the $C_6$ constant of the $^{40}$K$^{87}$Rb molecule and with an infinite $s$-wave scattering length.  The open and solid circles respectively show the calculated energy levels for the X$^1\Sigma^+$ and a$^3\Sigma^+$ potentials, for which the scattering lengths are $-111.8$ a$_0$ and $-216.2$ a$_0$ respectively ~\cite{Pashov2007}. The next lowest $n=-3$ levels of these potentials are at $-10.24$ GHz and $-10.56$ GHz.}
  \end{center}\label{Fig9}
\end{figure}

\clearpage

\begin{figure}[ht]
  \begin{center}
  \includegraphics[angle=270,width=\columnwidth]{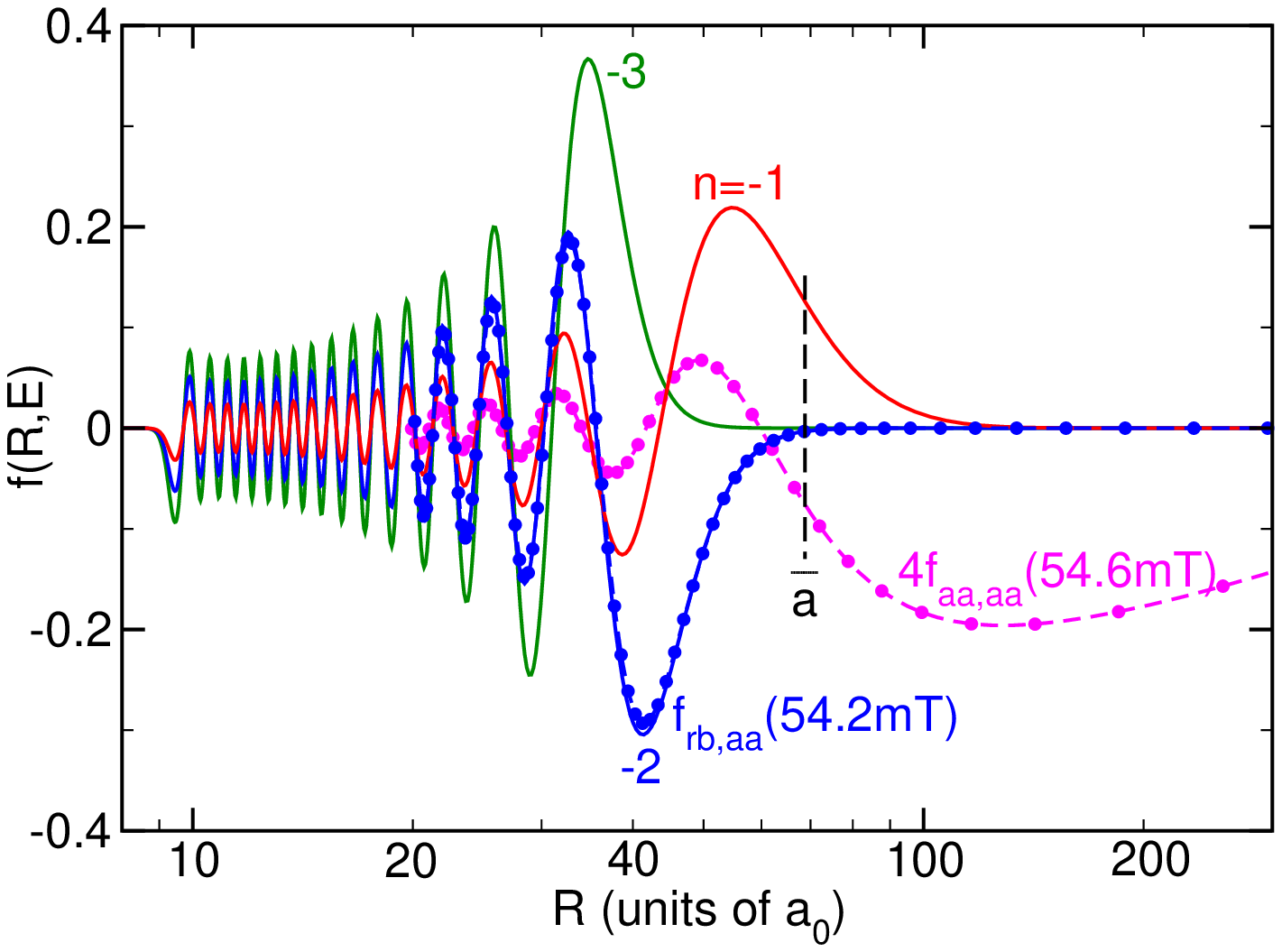}
   \caption{Calculated wave functions for near-threshold bound states of the $^{40}$K$^{87}$Rb molecule.  The solid lines labeled $n=-1,-2,-3$ show the unit-normalized wave functions $f_n(R)$ for the last three levels of the a$^3\Sigma^+$ potential.  The vertical line indicates the value of $\bar{a}$.  The dots and dashed lines show results from the coupled channels calculation of the $aa$ channel bound state of Eq.~\ref{eq:CC} near the 54.69 mT resonance, namely, the $rb$ component $f_{rb,aa}(R)$ at $B=54.2$ mT and the $aa$ channel component $f_{aa,aa}(R)$ at $B=54.6$ mT.  The latter represents a ''halo molecule'' with open channel $aa$ spin character and an extension large compared to $\bar{a}$.  The former represents a closed channel molecule with $rb$ spin character and a vibrational function very close to the $n=-2$ a$^3\Sigma^+$ vibrational wave function.}
  \end{center}\label{Fig10}
\end{figure}

\clearpage

\begin{figure}[ht]
  \begin{center}
  \includegraphics[angle=270,width=\columnwidth]{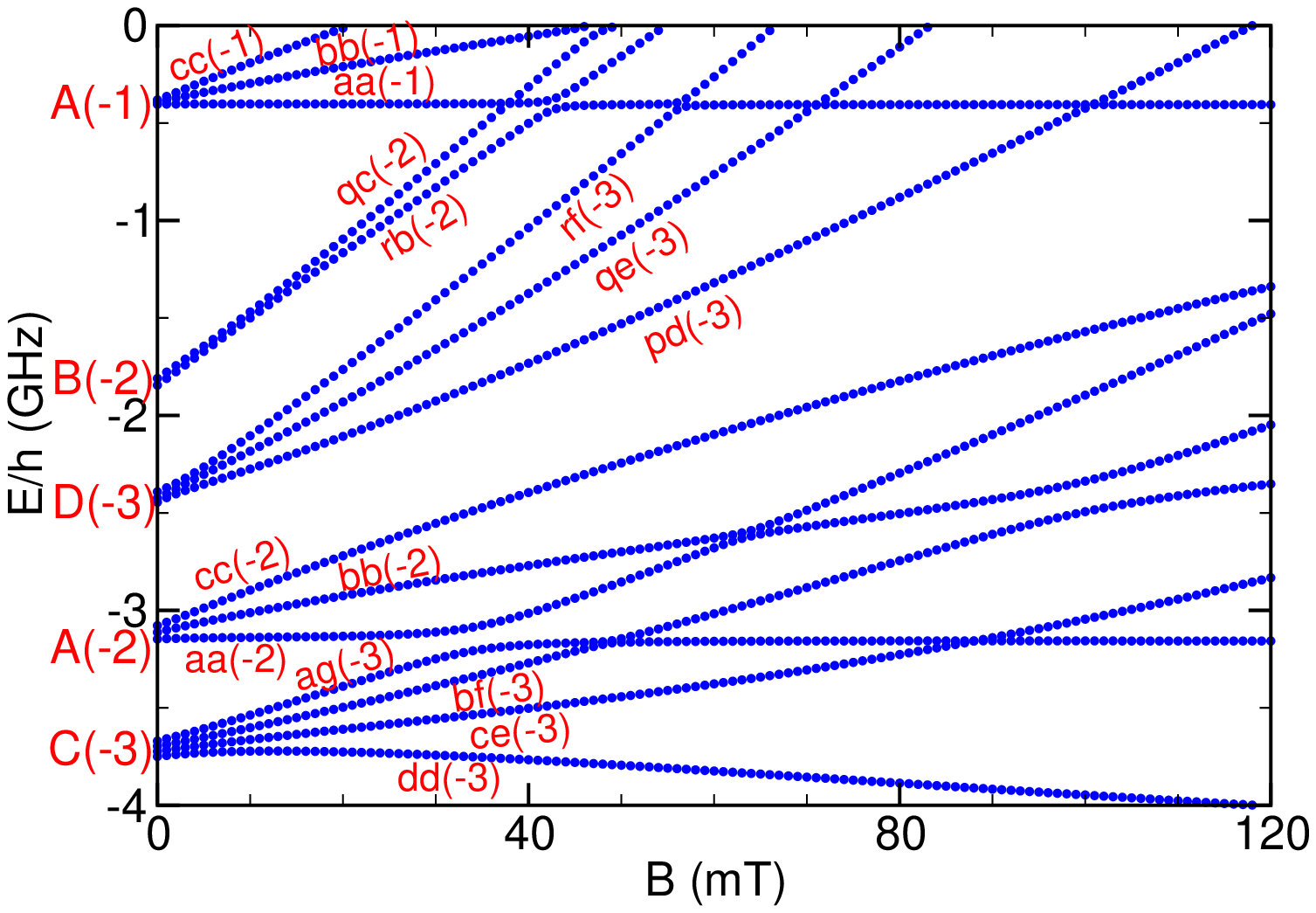}
   \caption{Bound state energy for the $M_\mathrm{tot}=-7/2$ $s$-wave bound states of the $^{40}$K$^{87}$Rb molecule down to 4 GHz binding energy.  The zero of energy is the energy $E_\alpha$ of the $aa$ channel at each $B$ field.  The level are labeled according to the channel index of their dominant spin component and by the vibrational quantum number $n$ counting down from the dissociation limit of the channel.  The Roman letters indicate the group in Figure 6 with which the $B=0$ channels are associated. }
  \end{center}\label{Fig11}
\end{figure}

\clearpage

\begin{figure}[ht]
  \begin{center}
  \includegraphics[angle=270,width=\columnwidth]{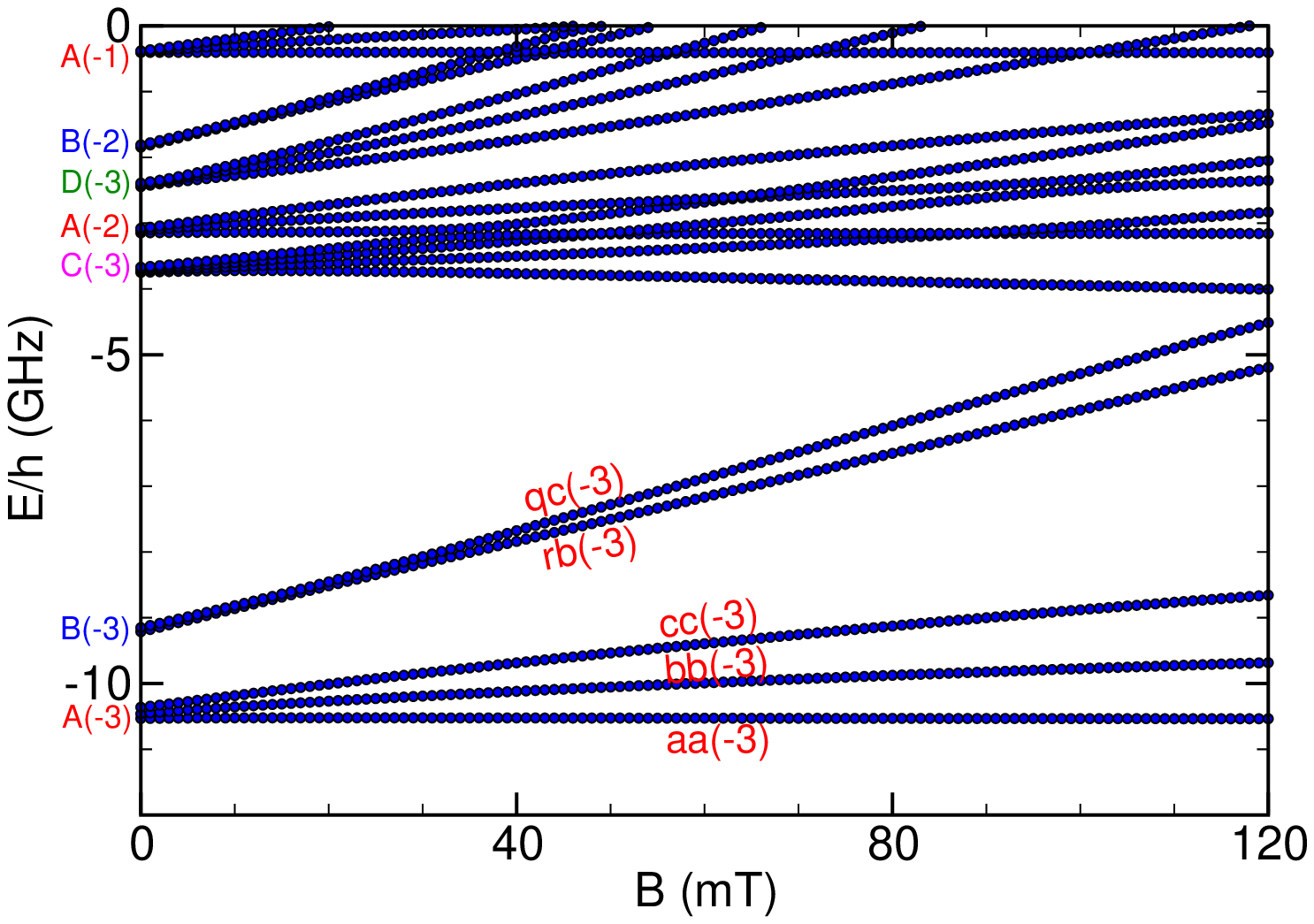}
   \caption{Bound state energy for the $M_\mathrm{tot}=-7/2$ $s$-wave bound states of the $^{40}$K$^{87}$Rb molecule down to 12 GHz binding energy.  The zero of energy is the energy $E_\alpha$ of the $aa$ channel at each $B$ field. Labels are the same as in Figure 11.}
  \end{center}\label{Fig12}
\end{figure}

\clearpage

\begin{figure}[ht]
  \begin{center}
  \includegraphics[angle=270,width=\columnwidth]{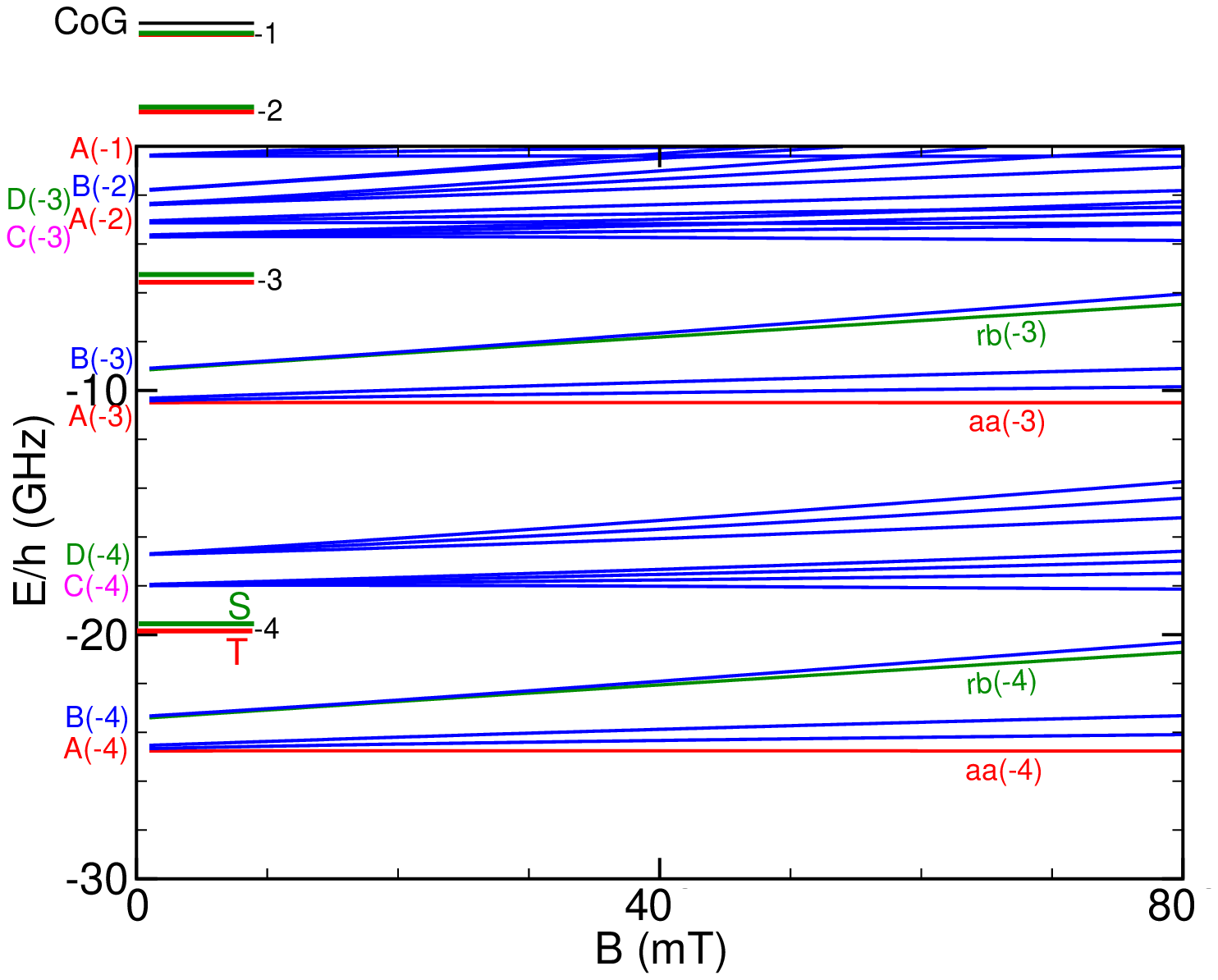}
   \caption{Bound state energy for the $M_\mathrm{tot}=-7/2$ $s$-wave bound states of the $^{40}$K$^{87}$Rb molecule down to 30 GHz binding energy.  The zero of energy is the energy $E_\alpha$ of the $aa$ channel at each $B$ field. Labels are the same as in Figure 11.  The short horizontal lines next to the $B=0$ axis labeled by vibrational quantum number $n$ show the energies of the levels of the X$^1\Sigma^+$ and a$^3\Sigma^+$  adiabatic Born-Oppenheimer potentials relative to the energy center of gravity (CoG) of the separated atom multiplet at $E/h=4.843439$ GHz. The X$^1\Sigma^+$ and a$^3\Sigma^+$ potentials support $N=$ 100 and 32 vibrational levels respectively.  The normal vibrational quantum number $v$ counting up from $v=0$ as the lowest level is $v=N+n$.}
  \end{center}\label{Fig13}
\end{figure}

\clearpage

\begin{figure}[ht]
  \begin{center}
  \includegraphics[angle=270,width=\columnwidth]{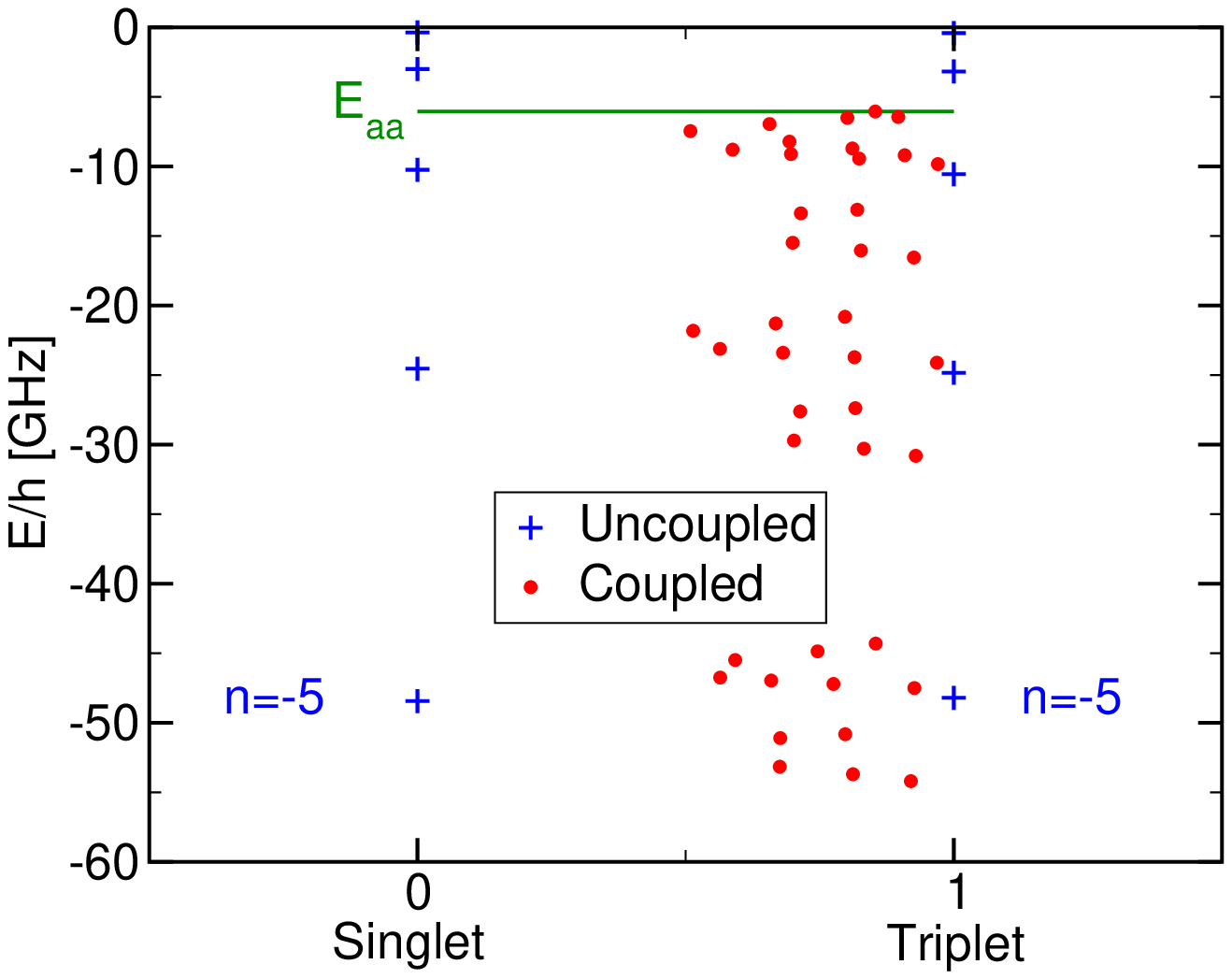}
   \caption{Bound state energy for the $M_\mathrm{tot}=-7/2$ $s$-wave bound states of the $^{40}$K$^{87}$Rb molecule at $B=54.6$ mT.  All energies are relative to the energy center of gravity of the atomic hyperfine multiplets.   The horizontal scale gives a measure of the singlet or triplet character of the eigenstate, with 0 representing a pure singlet state and 1 representing a pure triplet state.  The horizontal line at $E/h=-4.834$ GHz labeled $E_{aa}$ marks the energy of two $a$ state atoms on this scale.  The crosses labeled ''Uncoupled'' mark the energies of the X$^1\Sigma^+$ and a$^3\Sigma^+$ energy levels; $n=-5$ indicates the fifth level down in each potential.  The red dots labeled ''Coupled'' locate the actual eigenstates of the coupled channels calculation.  No levels have pure singlet character in this domain.}
  \end{center}\label{Fig14}
\end{figure}

\clearpage

\begin{figure}[ht]
  \begin{center}
  \includegraphics[angle=270,width=\columnwidth]{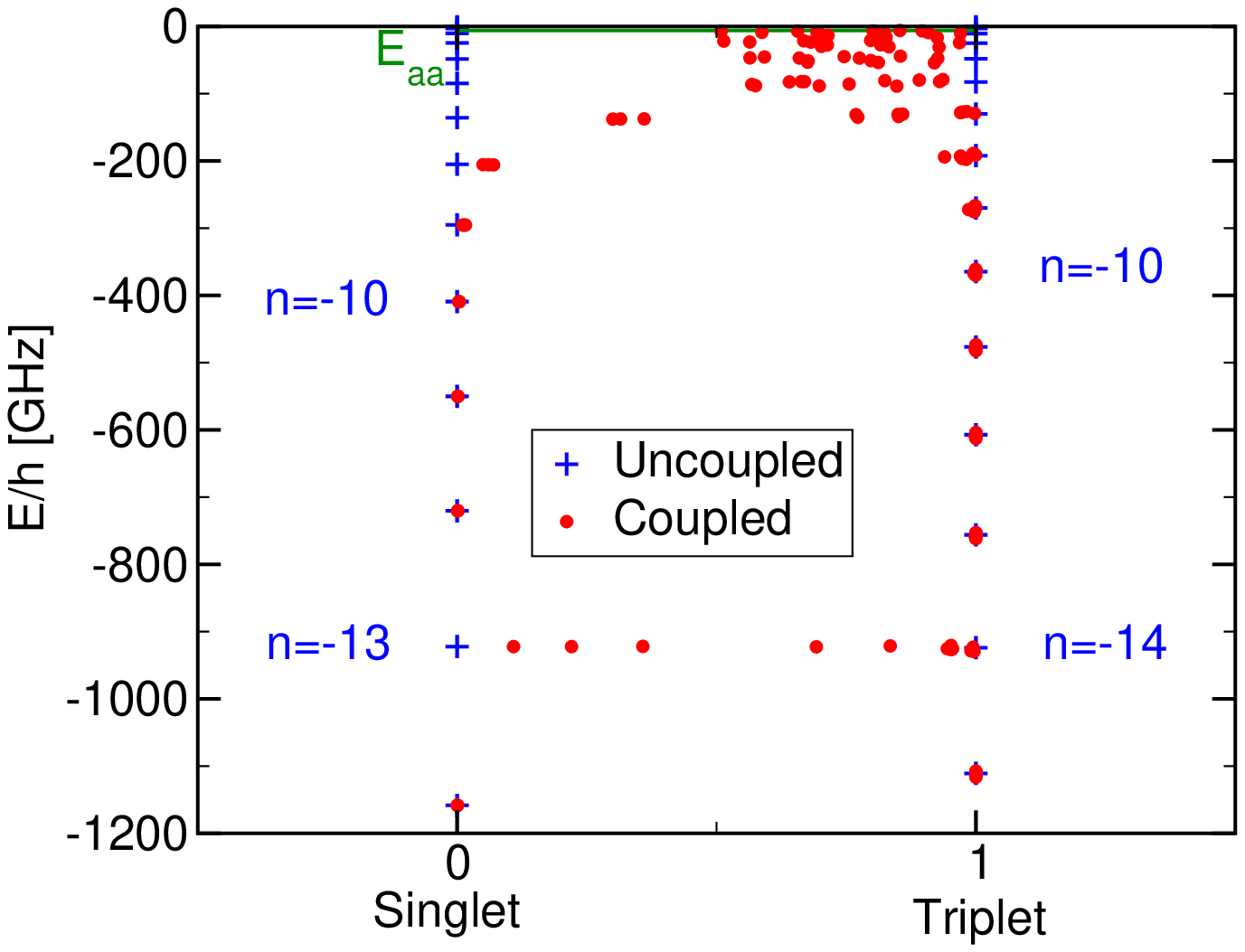}
   \caption{Bound state energy for the $M_\mathrm{tot}=-7/2$ $s$-wave bound states of the $^{40}$K$^{87}$Rb molecule at $B=54.6$ mT.  The labeling is the same as in Figure 14, but the energy scale extends to $E/h=1200$ GHz binding energy.  Uncoupled singlet levels with $|n| \le 7$ are strongly mixed with triplet levels of the same $n$.   The eigenstates separate into two sets of eigenstates of nearly pure singlet or triplet character when the binding energy becomes larger than around 200 GHz.  There is an accidental near-degeneracy of the $n=-13$ $(v=87)$ singlet level and $n=-14$ $(v=18)$ triplet level near 900 GHz binding that results in strong mixing between singlet and triplet levels in that energy range.  There are no more accidental degeneracies with strong perturbations for any  levels with lower energy. }
  \end{center}\label{Fig15}
\end{figure}

\clearpage

\begin{figure}[ht]
  \begin{center}
  \includegraphics[angle=270,width=\columnwidth]{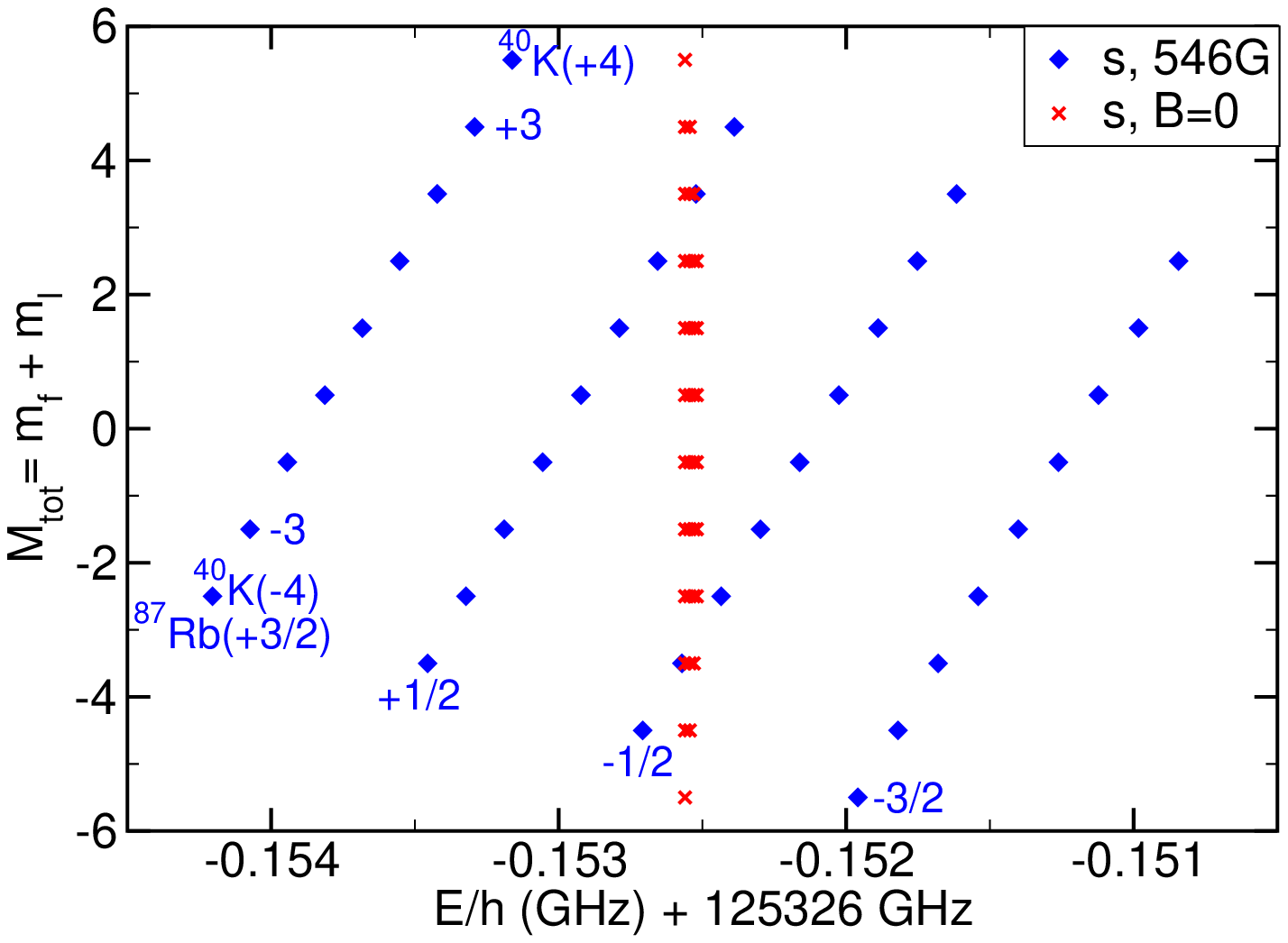}
   \caption{Bound state energy for all $M_\mathrm{tot}$ $v=0$ $J=0$ bound states of the $^{40}$K$^{87}$Rb molecule at $B=0$ (x symbols) and 54.6 mT (diamond symbols).   Energy is relative to the energy center of gravity of the atomic hyperfine multiplets.  At $B=0$ the levels divide into four degenerate groups with total nuclear spin quantum number $I_\mathrm{tot}=$ $11/2$, $9/2$, $7/2$, and $5/2$ in order of increasing energy, with a spread of 41 kHz between the highest and lowest energy.  At $B=$ 54.6 mT the nuclear spins become uncoupled from one another and represent two independent nuclear spins in the strong $B$ field.  The labels indicate the projection quantum numbers for the four projection levels of the $^{87}$Rb nuclear spin and the nine levels of the $^{40}$K nuclear spin.  }
  \end{center}\label{Fig16}
\end{figure}

\end{document}